\documentclass[twocolumn,twocolappendix]{aastex631}

\usepackage{amsmath}

\usepackage{orcidlink}
\newcommand{\AuthorORCID}[2]{%
  \href{https://orcid.org/#2}{\mbox{#1~{\Large\orcidlink{#2}}\kern-0.5em}}%
  }

\makeatletter
\def\switch@array{}
\makeatother
\usepackage{tabularx}

\begin{document}

\shorttitle{UHECR anisotropy favors a light composition}
\shortauthors{Strasman \& Waxman}

\title{The high entropy of the UHECR arrival direction distribution favors a light composition}

\author{\AuthorORCID{Nimrod Strasman}{0009-0008-8214-7677}}
\affiliation{Dept. of Particle Phys. \& Astrophys., Weizmann Institute of Science, Rehovot 76100, Israel}
\author{\AuthorORCID{Eli Waxman}{0000-0002-9038-5877}}
\affiliation{Dept. of Particle Phys. \& Astrophys., Weizmann Institute of Science, Rehovot 76100, Israel}

\begin{abstract}

We analyze the constraints on the composition of ultra-high-energy cosmic rays (UHECRs), and on the density and distribution of their sources, that may be inferred from their arrival-direction distribution, using a novel semi-analytical description of the propagation energy loss of $>2\times10^{19}~\text{eV}$ atomic nuclei ($A>4$), which yields insights into the impact of propagation energy loss. We show that the anisotropy of the UHECR arrival-direction distribution due to the large-scale structure (LSS) of the matter distribution is larger for heavy nuclei models (specifically, "Auger mix" models) than for proton composition models, despite larger magnetic deflections of nuclei, due to their smaller horizon and weaker dependence of rigidity on observed energy. Identifying the LSS signal is hampered for heavy nuclei by their large deflections by the uncertain Galactic magnetic field (GMF). We introduce a new measure of anisotropy, an "entropy" of the arrival-direction distribution, that is largely independent of the GMF and has discriminating power between heavy- and light-composition models. Analyzing the public $>3.2\times10^{19}$~eV Auger data, we show that the correlation with the LSS on large angular scales is weak and requires a low source density, $s_0\le10^{-4}{\rm Mpc}^{-3}$, to mask the LSS signature by "cosmic-variance" ($s_0=10^{-2}{\rm Mpc}^{-3}$ is ruled out at $>99\%$ confidence level (CL)). The high entropy of the distribution is consistent with proton models and inconsistent with heavy nuclei at $>96\%$ CL for $s_0\ge10^{-5}{\rm Mpc}^{-3}$. Reducing the energy calibration uncertainty may allow detection of the LSS correlation for proton models (increased exposure alone will not suffice due to cosmic variance).

\end{abstract}

\section{Introduction}
Identifying the sources of UHECRs and understanding their acceleration mechanisms are long-standing challenges of high-energy astrophysics \citep[see][for a recent review]{coleman_ultra_2023}. The arrival-direction distribution of cosmic rays carries information about their source distribution and composition, which, in turn, constrains source and acceleration models. While deflections of UHECRs, which are likely charged ions, by magnetic fields imply that their arrival directions do not point back directly to their sources, the correlation of the sources with the known LSS of matter in the local universe imprints an anisotropy signal that may be identified in the arrival direction distribution. For a light proton composition, the magnetic deflections are small compared to the characteristic angular anisotropy scale, allowing a direct identification of the LSS signature \citep{1997WaxmanFisher, abraham_correlation_2007, kashti_searching_2008, koers_testing_2009, takami_cross-correlation_2009, oikonomou_search_2013}. For a heavy nuclei composition, the deflections are large, and the uncertainties in the GMF structure \citep[e.g.,][]{korochkin_uhecr_2025} hamper such direct identification (the extra-Galactic magnetic field deflections are dominated by the GMF deflection, see \S~\ref{sec:magnet}). Nevertheless, as we show here in particular, some properties of the arrival-direction distribution are not sensitive to uncertainties in the GMF and allow discrimination between source and composition models.

Analyses of the Pierre Auger Observatory \citep[Auger;][]{pierre_auger_collaboration_pierre_2015} and the Telescope Array \citep[TA;][]{kawai_telescope_2008} data revealed several anisotropy signals.
\begin{itemize}
    \item A significant dipole anisotropy, $5.7\sigma$ at 8-16~EeV and 99\% CL at $>30$~EeV, had been observed in Auger data \citep{abdul_halim_large-scale_2024}, with no significant additional multipoles. The dipole amplitude is reproduced in simulations of UHECR propagation from the local universe \citep{bister_constraints_2024,allard_what_2026}, but its direction is not reproduced as easily, and depends on the GMF model used. These simulations assume a heavy UHECR composition and the picture might look different in a proton UHECR scenario. In the combined Auger + TA map the dipole is still significant and its direction seems to change dramatically with energy \citep{urena_new_2025}, although disagreements in the observed spectra of the two observatories \citep{the_telescope_array_collaboration_observation_2024} calls the validity of such combined analyses into question. The significant dipole is usually thought of as a signal of the LSS, but at lower UHECR energies of $\sim$\,10~EeV it might also be explainable by a galactic origin \citep{gruzinov_are_2018}.
    
    \item \citet{abreu_arrival_2022} observed a $4.2\sigma$ correlation with a sample of star-forming galaxies in Auger data, compared to weaker correlations with catalogs of different objects, e.g., active galactic nuclei (AGNs). Since the distributions of the different source types across the catalogs used in that work all follow the same LSS, these results hint at a correlation between UHECR sources and LSS but do not identify a particular source type. The distributions of different source types in the catalogs differ mainly in the sources' characteristic distances, which are determined by the catalogs' flux limits (for example, the distances of the star-forming galaxy sample are shorter than those of the AGN sample). Thus, the enhanced signal obtained for a correlation with star-forming galaxies provides information on the characteristic distance to the UHECR source, rather than on the source type \citep[consistent with the statement in][that their findings don't actually imply a firm preference to any of the catalogs]{abreu_arrival_2022}. The identified LSS correlation also favors a light, proton composition, since the analysis neglected deflections due to the coherent GMF, which would be large for heavy nuclei.
    
    \item Several small-scale excesses have been found in the data. Auger sees an excess in the Centaurus region with $4\sigma$ confidence \citep{pierre_auger_collaboration_update_2023}, TA sees an excess around the Perseus-Pisces region with $3.1\sigma$ confidence as well as the "TA hotspot" with $2.9\sigma$ \citep{kim_medium-scale_2025}. The Centaurus and Perseus-Pisces regions are notable overdensities in the LSS, while the TA hotspot does not correspond to any known structure. The significance of the TA hotspot decreased since its first detection, and no corresponding excess was found in Auger analysis of the same region of the sky \citep{abdul_halim_distribution_2025}, which calls its actual significance into question.
\end{itemize}

In this paper, we study constraints that may be inferred from the UHECR arrival-direction distribution on the distribution of UHECR sources and their composition, and that are robust with respect to uncertainties in the GMF structure. In \S~\ref{sec:analytic_treatment} we present a novel semi-analytical description of the propagation energy loss of atomic nuclei (\(A>4\)) with energy \(>2\times10^{19}~\text{eV} \), that allows generating UHECR arrival maps faster than by using detailed propagation simulations and yields insights to the characteristics of the impact of propagation energy loss. In \S~\ref{sec:magnet} we discuss deflections by extra-galactic and Galactic magnetic fields. In \S~\ref{sec:method} we derive the anisotropy imprinted by the LSS on the UHECR arrival direction distribution. We assume that the UHECR sources are steady and that their distribution follows the LSS with some bias, using the 2MASS Redshift Survey \citep[2MRS;][]{huchra_2mass_2012} to infer the LSS. We derive the expected arrival-direction distributions for the widely used heavy-nuclei composition UHECR models typically inferred from depth-of-shower-maximum analysis (the "Auger mix" models), as well as for a light proton composition.  
We introduce measures of anisotropy to identify LSS signatures and demonstrate the robustness of constraints obtained using them to GMF uncertainties by computing their distributions, accounting for deflections obtained for a wide range of models of the coherent GMF structure. In \S~\ref{sec:results} we use the methods developed in \S~\ref{sec:method} to analyze the public Auger data for $E>32~\text{EeV}$ and draw conclusions about the composition, the density of the sources and the bias of their distribution. We compare our results to those of earlier work in \S~\ref{sec:comparison}. Our conclusions are summarized and discussed in \S~\ref{sec:discussion}. In particular, we discuss how the methods derived here can be used to infer improved constraints using future data. A code implementing these methods is available at \url{https://github.com/Nimrod-S/crtools}.

Throughout the paper we use a flat \(\Lambda\text{CDM}\) universe with \(\Omega_m =0.3,H_0=70~\text{(km/s)/Mpc}\).

\section{Analytic treatment of propagation} \label{sec:analytic_treatment}
In the following sections we describe how we model the energy loss of nuclei and the propagation of source spectra. When proton UHECRs are considered, we use the energy loss calculation method outlined in \citet{katz_energy_2009} with a volumetric source spectrum of 
\begin{equation}\label{eq:protonspectrum}
    J(E)=3.53\times10^{43}\left(\frac{E}{10^{19.6}~\text{eV}}\right)^{-0.5}~\text{rays}~\text{eV}^{-1}\text{Mpc}^{-3}\text{yr}^{-1}.
\end{equation}

\subsection{Energy loss}
At the relevant energies (\(E\gtrsim2\times10^{19}~\text{eV}\)), energy loss of nuclei is dominated by photodisintegration, which reduces the mass \(A\) and therefore the energy \(Am_p\gamma c^2\) of the nucleus, and pair production, which instead affects the Lorentz factor \(\gamma\) \citep[e.g.,][]{allard_extragalactic_2012}. We are going to neglect processes of the second kind, which have a comparatively smaller effect, and model solely the effect of photodisintegration. We do take into account the effect of the expansion of the universe on the energy (and hence on the cross sections) and on the photon background density. We ultimately justify these assumptions by checking our results against a more accurate simulation.

For \(A\gg 4\), we approximate the mass loss as a continuous process, described as 
\begin{equation}
    \frac{dA}{dz}=-\frac{dx}{dz}\sum_i\frac{n_i}{\lambda_i(A,\gamma, z)},
\end{equation}
where \(A\) is the mass number, \(z\) is redshift, \(x\) is proper distance, the sum is over different photodisintegration processes that each eject a different number of nucleons \(n_i\) from the nucleus, and \(\lambda_i \) are the mean free paths (MFPs) for each of them. We define the effective MFP (henceforth written as \(\lambda\) with no lower index) as
\begin{equation}
\lambda^{-1}=\sum_in_i\lambda_i^{-1},
\end{equation}
which simplifies the expression to
\begin{equation}
    \frac{dA}{dz}=-\frac{dx}{dz}\frac{1}{\lambda(A,\gamma, z)}.
\end{equation}

Note that, since processes that eject a single nucleon are almost always dominant, it is roughly true that \(\lambda\approx\lambda_1\), but there is no reason to make this approximation.

It is technically not correct that \(\lambda\) depends on the mass \(A\) alone, since the cross sections should also depend on the charge. Still, for every mass \(A\) we are going to only look at the stable isotope with that mass under the assumption that unstable isotopes either decay quickly or have a similar cross section to the corresponding stable isotope with the same mass. This assumption too will be justified by comparing our final results to a full simulation.

\begin{figure*}
    \centering
    \includegraphics[width=1\linewidth]{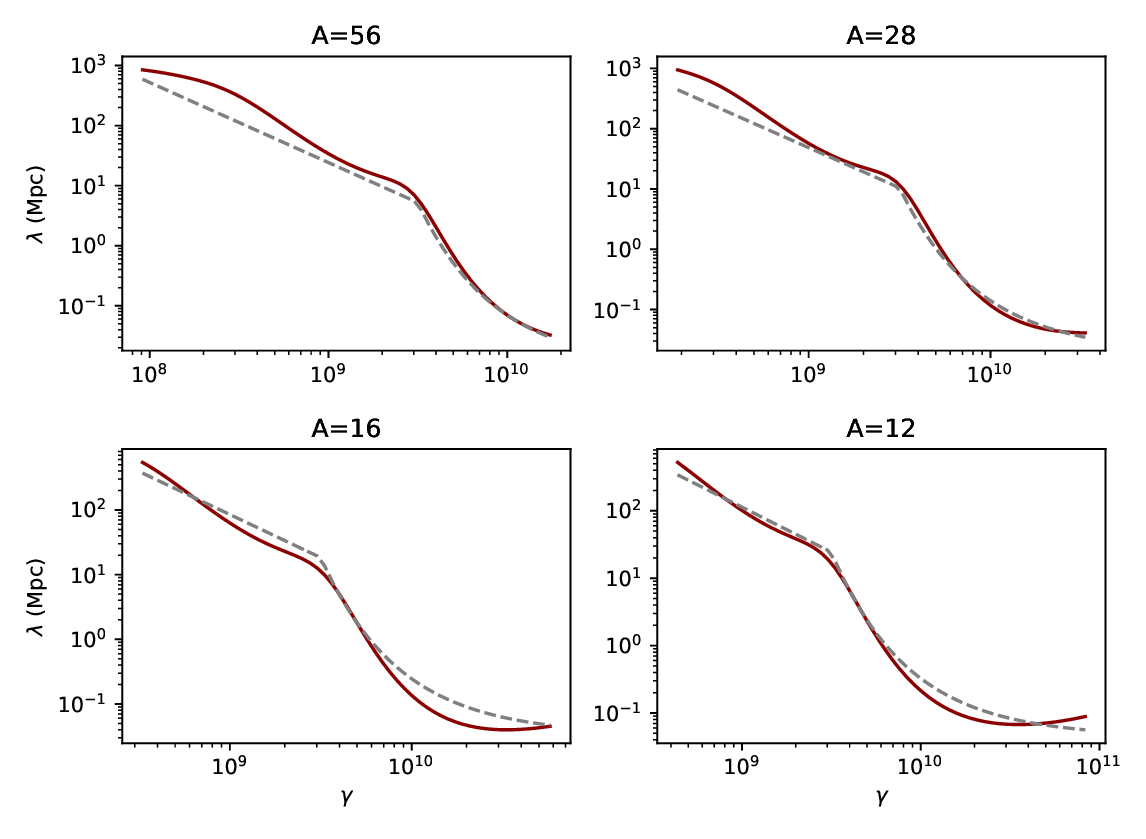}
    \caption{Comparison of the approximate effective MFPs (dashed) to the more precise values (full) obtained from the photodisintegration tables of CRPropa3 \citep{batista_crpropa_2022}, using the Gilmore12 model for the IRB \citep{gilmore_semi-analytic_2012}, for Fe, Si, C and O nuclei.}
    \label{fig:mfp_comp}
\end{figure*}

The values of the photodisintegration MFPs at redshift \(z=0\) are approximated as follows:
\begin{equation}
    \lambda(A,\gamma)\approx A^{-1}\lambda(\gamma),
\end{equation}
\begin{equation}\label{eq:energyloss}
    \lambda(\gamma)\approx C\times\begin{cases}
        e^{\gamma_c/\gamma_b}(\gamma/\gamma_b)^{-k}&\gamma<\gamma_b\\
                e^{\gamma_c/\gamma}&\gamma>\gamma_b\\
    \end{cases}.
\end{equation}
\(C,\gamma_c,\gamma_b,k\) are given in Table~\ref{tab:mfp_params}. \(\gamma_b\) represents the transition between interaction with the infrared background (IRB) and the CMB. \(\gamma_c\) controls the shape in the CMB regime, \(k\) controls the shape in the IRB regime, and \(C\) is some distance scale. As we can see in Fig.~\ref{fig:mfp_comp}, the model is generally in good agreement with the exact values. It is the least accurate for iron, and we will see that our energy loss prediction for iron is indeed the worst, but this won't meaningfully affect the final results.

\begin{table}
    \centering
    \begin{tabularx}{\linewidth}{*4{>{\centering\arraybackslash}X}}
        \hline
        $C$ & $\gamma_b$ & $\gamma_c$ & $k$ \\
        \hline
        $0.532$~Mpc& $10^{9.5}$ & $10^{10.3}$ & $4/3$\\
        \hline
    \end{tabularx}
    \caption{Values used in energy loss modeling. See Eq.~(\ref{eq:energyloss}).}
    \label{tab:mfp_params}
\end{table}

Finally, the dependence of \(\lambda\) on redshift will be approximated as
\begin{equation}
    \lambda(A,\gamma,z)\approx(1+z)^{-(3+2k)}\lambda(A,\gamma).
\end{equation}

The factor \((1+z)^{-3}\) is due to the increasing density of the photon fields as we go back in time which directly shrinks the MFP. The factor \((1+z)^{-2k}\) is due to both the increasing energy of the IRB and the increasing Lorentz factor of the ray itself. This factor is only correct in the IRB interaction regime, but we use it for every \(\gamma\) to make the calculation simpler. Since the CMB interaction lengths are rather short, rays whose propagation is dominated by CMB interactions must come from nearby where redshift effects are negligible anyway.

We can now solve the differential equation by integrating the mass loss from the value at the source \(A_s\) to the observed value \(A_o\):
\begin{equation}
    \frac{dA}{dz}=-\frac{dx}{dz}(1+z)^{3+2k}A\lambda(\gamma)^{-1},
\end{equation}
\begin{equation}
    \int_{A_o}^{A_s} A^{-1}dA=\lambda(\gamma)^{-1}\int_0^zc\frac{dt}{dz}(1+z)^{3+2k}dz.
\end{equation}

\begin{figure}
    \centering
    \includegraphics[width=1\linewidth]{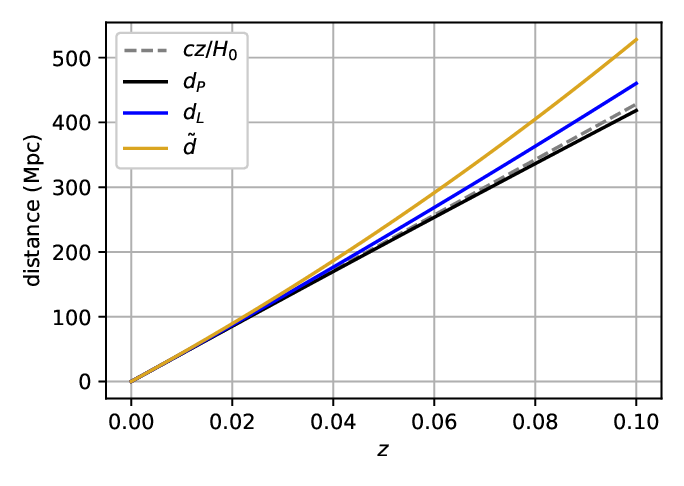}
    \caption{Comparison of the effective propagation distance \(\tilde d\) to luminosity distance, proper distance and the naive Hubble's law.}
    \label{fig:eff_dist}
\end{figure}

We will name the integral over redshift the "effective propagation distance"
\begin{equation}
    \tilde d(z)= \int_0^zc\frac{dt}{dz}(1+z)^{3+2k}dz
\end{equation}
and calculate it numerically when necessary. It is a similar calculation to the proper distance \(d_P(z)=\int_0^zc\frac{dt}{dz}(1+z)dz\). A comparison to other distance measures is given in Fig.~\ref{fig:eff_dist}.

The solution ends up being
\begin{equation}
    \ln{A_s}-\ln{A_o}=\lambda(\gamma)^{-1}\tilde d(z),
\end{equation}
\begin{equation}
    A_o=A_s\times e^{-\tilde d(z)/\lambda(\gamma)},
\end{equation}
and, since under our assumptions the Lorentz factor does not change in propagation, this exponential factor will also be the energy loss factor
\begin{equation}
    E_o=E_s\times e^{-\tilde d(z)/\lambda(\gamma)}.
\end{equation}

\begin{figure*}
    \centering
    \includegraphics[width=1\linewidth]{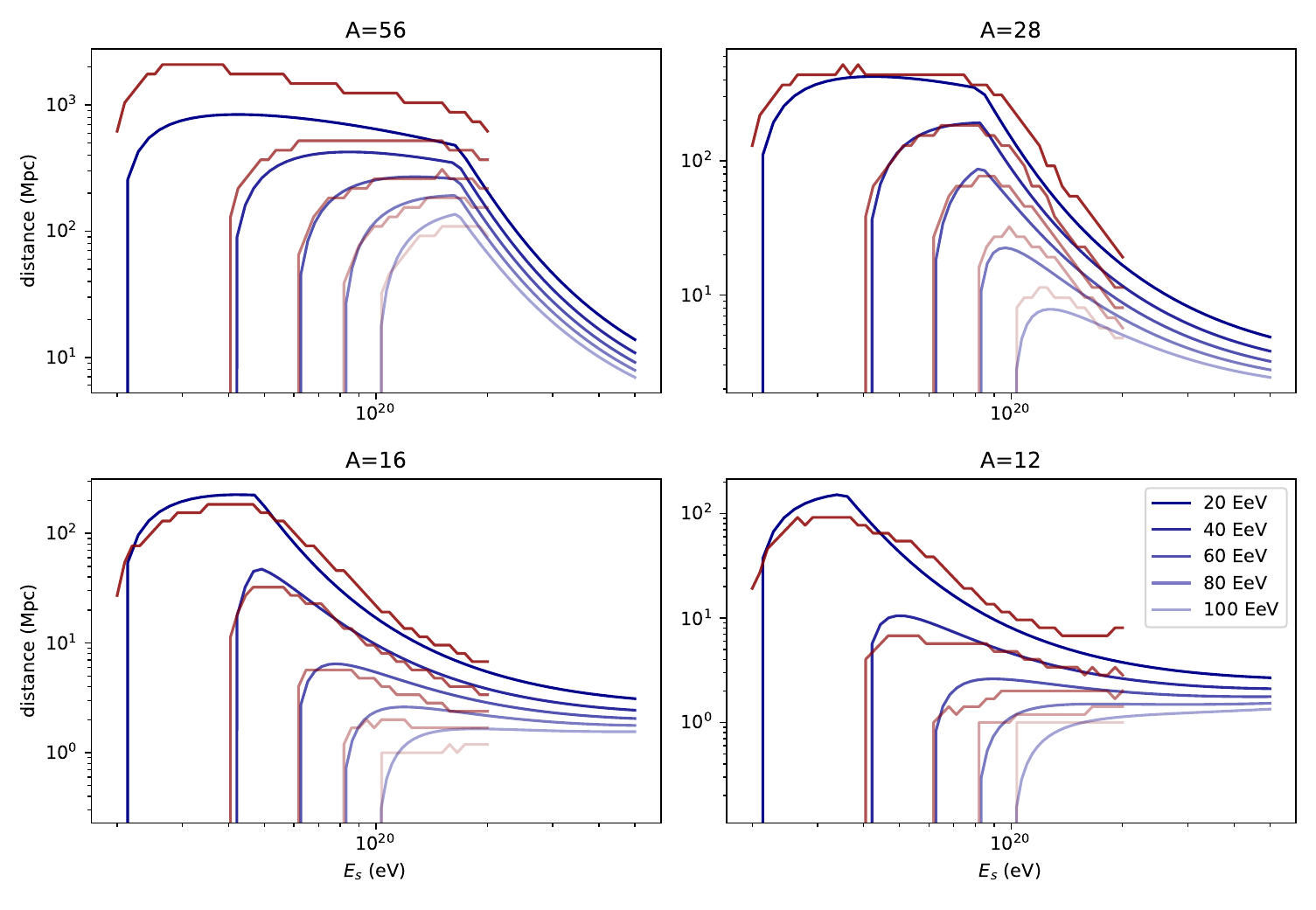}
    \caption{The connection between energy at the source \(E_s\) and the distance from the source, for different observed energies \(E_o\) (see legend), for Fe, Si, C and O nuclei. The effective distances are calculated with Eq.~(\ref{eq:dtilde_by_E}), then converted to the more meaningful proper distance for this graph. Our model (blue) is compared to the results of CRpropa3 simulations (red). Our approximations are not good for iron at low energies, but in the source models we are using, there is very little iron contribution below $4\times10^{19}$~eV.}
    \label{fig:econtours}
\end{figure*}

With this relatively simple expression for energy loss, for a given observed comic ray energy \(E_o\), we can find a connection between the possible source distances \(\tilde d\) and source energies \(E_s\):
\begin{equation} \label{eq:dtilde_by_E}
    \tilde d = \ln{(E_s/E_o)}\lambda(E_s/(A_sm_pc^2))
\end{equation}
as plotted in Fig.~\ref{fig:econtours}.

This expression can also be written as 
\begin{equation}
\tilde d =\ln{(E_s/E_o)}A\lambda(A,E_s/(A_sm_pc^2)),
\end{equation}
so instead of depending on our approximation for \(\lambda(\gamma)\), we can instead directly plug in the known $\lambda(A,\gamma)$ values. This is what we will do when we use \(\tilde d \) for the cosmic ray propagation calculations later on.

Finally, to check the validity of all our approximations, we compare our results to more precise propagation simulations using CRPropa3 \citep{batista_crpropa_2022}, also in Fig.~\ref{fig:econtours}. For every source energy \(E_s\) we inject rays from farther and farther distances until the mean observed energy for a 1000 simulated rays is equal to \(E_o\) to plot the red contours. The worst match is for iron at low energies, but we are not particularly worried about it since in the source spectrum models we are using, iron is dominant only at the highest energies. The good agreement otherwise justifies some of the less obvious approximations like neglecting pair-production energy loss and the treatment of redshift, at least when considering rays with \(E>2\times10^{19}~\text{eV}\) (keep in mind that since there are more rays with lower energies than with higher energies, we will not be very sensitive to the details of the relatively few \(\gtrsim8\times10^{19}~\text{eV}\) rays).

One interesting observation is that, for nuclei, there exists a maximum, "hard limit" on the distance that the UHECRs could arrive from. This is unlike protons, since the proton energy loss processes depend on \(\gamma\) alone and the arrival energy is monotonously rising with the source energy, so theoretically protons could arrive from arbitrarily far (of course limited by the finite energy of the acceleration mechanism). On the other hand for nuclei, if the Lorentz factor gets too high, the photodisintegration becomes too efficient and reduces the energy faster than it would for a less energetic nucleus. By differentiating equation Eq.~(\ref{eq:dtilde_by_E}) we find the maximum distance:

\begin{equation}
    \lambda^\prime(\gamma)=\begin{cases}
    -\frac{k}{\gamma}\lambda(\gamma)&\gamma<\gamma_b\\
    -\frac{\gamma_c}{\gamma^2}\lambda(\gamma)&\gamma>\gamma_b
    \end{cases},
\end{equation}
\begin{align}
    \frac{d\tilde d}{d\gamma}&=\frac{1}{\gamma}\lambda(\gamma)+\ln(E_s/E_0 )\lambda^\prime(\gamma)\\
    &=\frac{1}{\gamma}\lambda(\gamma)\begin{cases}
        1-k\ln(A_sm_pc^2\gamma/E_o)&\gamma<\gamma_b\\
        1-\frac{\gamma_c}{\gamma}\ln(A_sm_pc^2\gamma/E_o)&\gamma>\gamma_b\\
    \end{cases}.
\end{align}

The low energy case is easily solved for, \begin{equation}
    d_m=\lambda(\gamma_m)/k,\gamma_m=\frac{E_o}{A_sm_pc^2}e^{1/k},
\end{equation} although this is valid only when \(\gamma_m<\gamma_b\). In the high energy case \begin{equation}
    d_m=\lambda(\gamma_m)\gamma_m/\gamma_c,
\end{equation} where \(\gamma_m\) is the solution of \begin{equation}
    \gamma_m=\gamma_c\ln(A_sm_pc^2\gamma_m/E_o),
\end{equation} valid when \(\gamma_m>\gamma_b\). Finally, there is also the in-between case where \begin{equation}
    d_m=\lambda(\gamma_b)\ln(A_sm_pc^2\gamma_b/E_o),\gamma_m=\gamma_b.
\end{equation}

\subsection{Propagating the source spectra}\label{sec:spectrum}

Ultimately we would like to reproduce the observed UHECR spectrum using our approximation. For this we need to know the emission spectrum of the sources, which we will take to be identical for all sources.

Heavy nuclei models of UHECRs usually assume a spectrum of the form
\begin{equation}\label{eq:theirspectrum}
    \frac{dQ_A}{dE}=J_A(E)=f_AJ_0E^{-\gamma_{ind}}e^{-E/Z/R_c}
\end{equation}
for each nucleus \(A\), where \(R_c\) is some cutoff rigidity above which the source can't accelerate the rays effectively and \(f_A\) is a fraction of this nucleus of the total. \(J_A\) measures number of rays per unit of energy, volume and time.

This way of writing the spectrum is not ideal since \(f_A\) has no real physical meaning. Since an electromagnetic process will accelerate nuclei with the same rigidity similarly, it is more meaningful to look at the nuclei fractions for a given rigidity
\begin{equation}\label{eq:ourspectrum}
    \frac{dQ_A}{dR}=J_1\tilde f_{A}\left(\frac{R}{R_c}\right)^{-\gamma_{ind}}e^{-R/R_c},
\end{equation}
where \(J_1=J_0R_c^{-\gamma_{ind}}\), \(\tilde f_{A}=f_AZ^{1-\gamma_{ind}}\) and the total volumetric energy luminosity summed over the entire spectrum is \(Q=\Gamma(2-\gamma)J_1R_c^{2}\sum_A\tilde f_{A}Z\). These \(\tilde f\)s are the "real" fractions of the different species at the source. The exact values of the parameters we are using in this work are presented in Table~\ref{tab:spec_params}, and were adopted from one of the models in \citet{abdul_halim_constraining_2024}, then accommodated a little to account for rounding off the power law to \(\gamma_{ind}=-2\). Overall these are typical values for these UHECR nuclei models that assume a single type of sources. We are not actually calculating the contribution of He to the spectrum as it is only relevant at lower energies, but its fraction is still important for the overall normalization.

\begin{table*}
    \centering
    \begin{tabularx}{\linewidth}{*7{>{\centering\arraybackslash}X}}
        \hline
        $\gamma_{ind}$ & $R_c~(\text{V})$ & $\tilde f_{\text{He}}$ & $\tilde f_{\text{N}}$ & $\tilde f_{\text{Si}}$ & $\tilde f_{\text{Fe}}$ & $Q~(\text{erg}~\text{Mpc}^{-3}~\text{yr}^{-1})$ \\
        \hline
        $-2$& $10^{18.15}$ & $1$ & $0.886$ & $0.064$ & $0.006$ & $5.85\times10^{44}$\\
        \hline
    \end{tabularx}
    \caption{Source emission parameters used in this paper for the nuclei UHECR model. See Eq.~(\ref{eq:ourspectrum}).}
    \label{tab:spec_params}
\end{table*}

\begin{figure}
    \centering
    \includegraphics[width=1\linewidth]{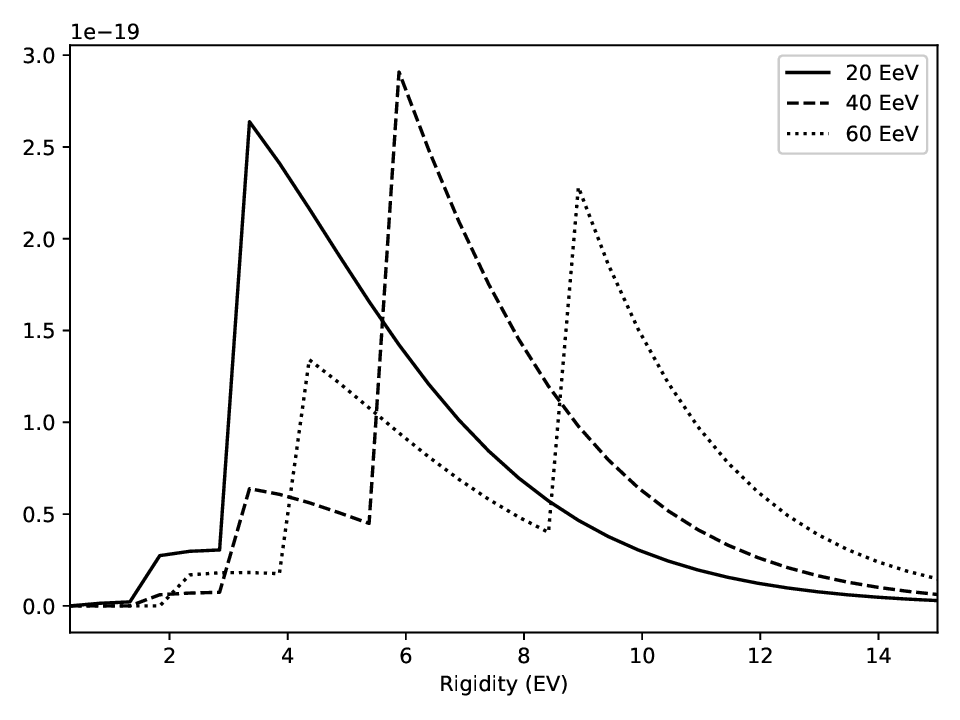}
    \caption{The (normalized) rigidity distribution of rays above given observed energies.}
    \label{fig:rigidity}
\end{figure}

\begin{figure*}
    \centering
    \includegraphics[width=1\linewidth]{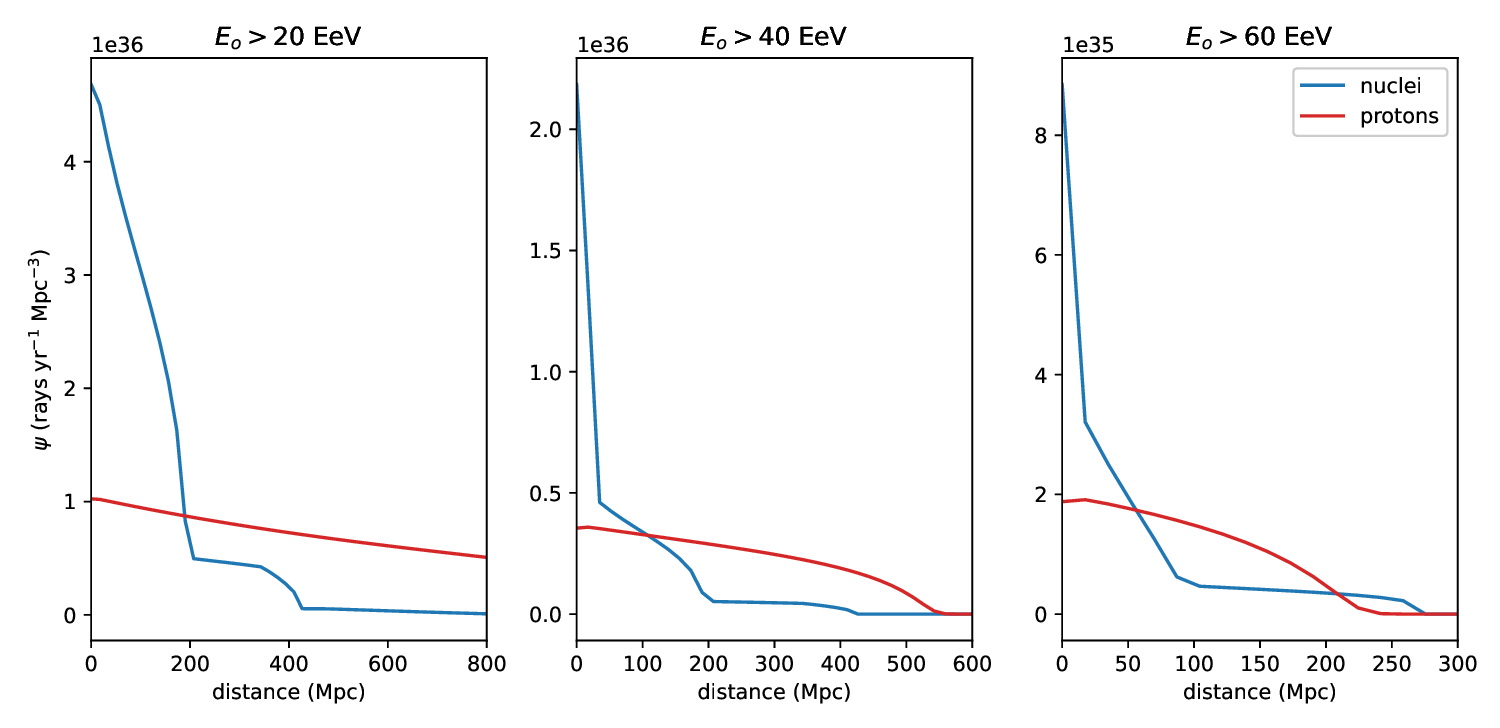}
    \caption{The volumetric UHECR emission rate \textit{of rays arriving above $E_o$} at different distances, for rays above some observed energies, described by Eq.~(\ref{eq:psi_dndr}). The results for the heavy nuclei model are compared to the pure proton model.}
    \label{fig:dndr}
\end{figure*}

To get the total number flux on Earth above some energy \(E_o\), we need to integrate \(J(E)\) over the entire spatial volume and energy range that will produce a ray above \(E_o\) on Earth. This is equivalent to integrating over the area under the corresponding \(E_o\) contour as seen in Fig.~\ref{fig:econtours}, since the contour tells us which distances and source energies will give us a ray with exactly \(E_o\) on Earth. There are two ways of calculating this integral; either first integrating "vertically" (over spatial volume), then integrating "horizontally" (over the source energy spectrum), or the other way around. The intermediate results for both of these methods are interesting by themselves:
\begin{itemize}
    \item In the first method, the result of the first integration is
    \begin{equation}
        \frac{d\phi(E_s;E_o)}{dE_s}=\sum_A\int_0^{\tilde d(E_o,E_s,A)} d\tilde d\frac{dr}{d\tilde d}J_A(E_s)(1+z)^{-3},
    \end{equation}
    which is approximately
    \begin{equation}
        \frac{d\phi(E_s;E_o)}{dE_s}\approx\sum_Ad(E_o,E_s,A) J_A(E_s),
    \end{equation}
    where \(d\) is the proper distance associated with \(\tilde d\). This allows us to calculate the rigidity distribution at the source, and assuming the rigidity of the rays does not change significantly during propagation (since energy loss is dominated by mass loss, and the mass/charge ratio remains similar), this would also be the rigidity distribution at arrival:
    \begin{equation} \label{eq:rigdist}
        \frac{d\phi(R;E_o)}{dR}=\sum_A d(E_o,RZ,A)J_A(RZ)Z.
    \end{equation}
    
    This is a particularly important result since the rigidity controls the magnetic deflections. A few distributions are shown in Fig.~\ref{fig:rigidity}. In the figure one can clearly see that as the energy goes up, the N component shifts to higher rigidities, but the Si component emerges in the lower end, so overall the distribution ends up moving very slowly to the right. This is a feature of this source emission model where the sources accelerate rays to a narrow rigidity range.
    \item In the second method, the result of the first integration is
    \begin{equation} \label{eq:psi_dndr}
        \psi(\tilde d;E_o)=\sum_A\int_{E_{min}}^{E_{max}} dE J_A(E),
    \end{equation}
    where \(E_{min},E_{max}\) are the points where the horizontal \(\tilde d\) line crosses the \(E_o\) contour (which we find numerically). Note that for some \(\gamma_{ind}\) values, in our case \(-2\), this integral is solvable analytically.

    This quantity \(\psi\), the rate of rays above arrival energy \(E_o\) emitted from a volume element, will serve us in the next sections. It also allows us to see the distance distribution from which we receive rays in some energy range, visualized in Fig.~\ref{fig:dndr}. The fact that the nuclei UHECR sources must be much closer than proton sources is immediately apparent.
\end{itemize}

\begin{figure}
    \centering
    \includegraphics[width=1\linewidth]{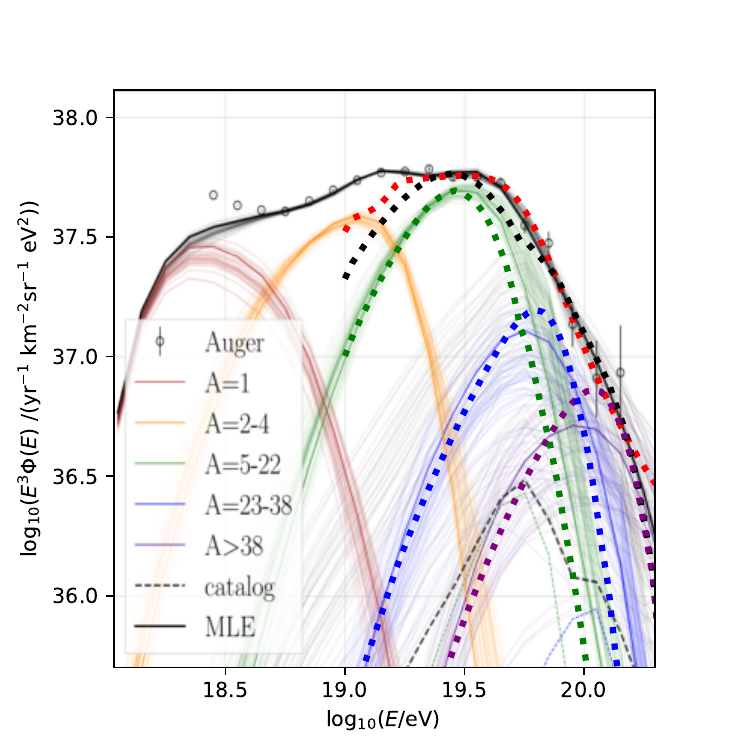}
    \caption{Our results (dashed lines) for the arrival spectrum in a heavy UHECR scenario compared to one of the models in \citet{abdul_halim_constraining_2024} (solid lines, graphs taken from their paper). The black line is the total spectrum for the heavy-composition model, and the purple, blue, and green lines correspond to the spectra of different mass ranges. Auger data are shown by empty circles, and our proton UHECR model is shown by the red dashed line. Note that since our energy loss calculations aren't applicable for nuclei with \(A<4\), we do not include the He component of the nuclei spectrum. We note that the \citet{abdul_halim_constraining_2024} model in this plot includes both the contribution from a source distribution following the LSS and an additional contribution from a nearby source, Centaurus A. However, the fractional contribution of this source to the total flux shown is small, $\simeq2.8\%$.}
    \label{fig:spectrum}
\end{figure}

\begin{figure}
    \centering
    \includegraphics[width=1\linewidth]{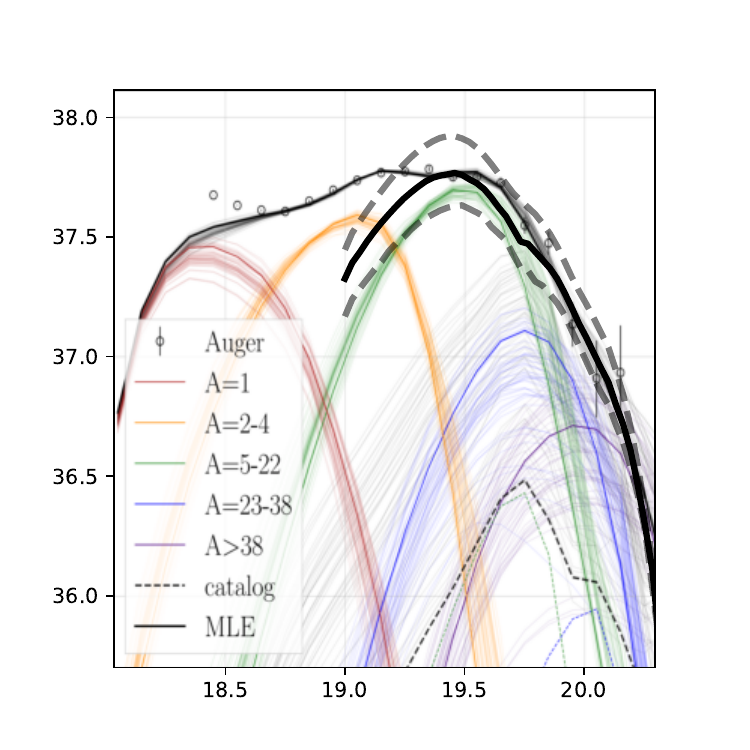}
    \caption{Similar to Fig.~\ref{fig:spectrum}, showing only the full spectrum of nuclei. The solid black line is obtained using the Gilmore12 model for the IRB \citep{gilmore_semi-analytic_2012}, used in the rest of our work. The dashed gray lines are obtained using the upper and lower limit Stecker16 models \citep{stecker_empirical_2016}, which reflect observational uncertainties.}
    \label{fig:spectrum_irb}
\end{figure}

Both results can then be integrated further, then differentiated by \(E_o\) to give us the spectrum. Fig.~\ref{fig:spectrum} shows the results, compared to one of the models in \citet{abdul_halim_constraining_2024}. Under the full spectrum are the "partial" spectra of rays that arrive in specific mass bins.
The good agreement of the spectrum above \(2\times10^{19}~\text{eV}\), up to about \(20\%\) at worst, justifies the use of our semi-analytic method for the analysis in this paper. The deviation of the analytic result from the numeric spectrum is smaller than the uncertainty arising from IRB uncertainties, as can be seen in Fig.~\ref{fig:spectrum_irb}, which is in principle an uncertainty that all propagation models suffer from.


\section{Magnetic deflections} \label{sec:magnet}
\subsection{GMF}

The biggest uncertainties in modeling UHECR propagation arise from GMF model uncertainties. In this work, rather than committing to a single model, we focus on observables that are not sensitive to the exact GMF model used.

\subsubsection{The coherent component}
We use the UF23 suite of 8 models presented in \citet{unger_coherent_2024} and, in a few cases, also the KST24 model \citep{korochkin_coherent_2025} as a basis for assessing uncertainties.
For each arrival energy bin $E_i$, we calculate the mean rigidity of the arriving rays (using Eq.~(\ref{eq:rigdist}) for nuclei). Then for each such rigidity and each model $m=1,2,\dots8$ in the UF23 list, we backtrack rays using CRPropa to generate a deflection lens $L_{m,i}(l,b)$ (where $m$ indexes the GMF model and $i$ indexes the energy bin), which is a mapping from an arrival direction on Earth $l,b$ to an arrival direction to the galaxy. These lenses will be used later when we simulate UHECR arrival maps.

Coherent deflections can be quite large, at least in the case of a nuclei UHECR scenario, and make it difficult to correlate the arrival map with specific structures in the local universe. One way to deal with it is to only correlate the UHECR map with the LSS on very large angular scales. Another is to take advantage of the fact that while the coherent field can move anisotropies around, it can only wash them out via demagnification, so the variance of the intensity distribution across the sky is still a good indicator of anisotropy. We take both of these approaches in this work when we design our tests.

\subsubsection{The random component}

The RMS of the deflection angle of a particle propagating a distance $D$ through a field with correlation $l$ and Larmor radius $R_L\gg l$ is \citep{waxman_images_1996} \begin{equation} \label{eq:magdef}
    \theta =\sqrt{\frac{2}{9}}\frac{\sqrt{Dl}}{R_L}.
\end{equation}

As a rough estimation, taking $B=5~\mu\text{G},\lambda=55~\text{pc}$ \citep{beck_magnetic_2016} and assuming a path of $2~\text{kpc}$ through the thick disc of the Milky Way, we expect UHECR nuclei deflections to be on the order of 
\begin{equation}\label{eq:basic_random_deflection}
    \theta\approx13.8^{\circ}\left(\frac{R}{3~\text{EV}}\right)^{-1}\left(\frac{D}{2~\text{kpc}}\right)^{1/2},
\end{equation}
which is not negligible but is not big enough to completely erase anisotropy signals.

An alternative value for the random deflection angle can be inferred from RM measurements. The relation between the standard deviation of RM measurements over some part of the sky to the deflection angle is  
\begin{eqnarray} 
    \sqrt{\langle\theta^2\rangle}=&\frac{1}{\sqrt{3}}&2.8\times10^3\left(\frac{n_e}{0.01~\text{cm}^{-3}}\right)^{-1} \nonumber\\
    &\times&\left(\frac{R}{40~\text{EV}}\right)^{-1}\sqrt{\langle 
    \text{RM}^2\rangle~\text{m}^4},
\end{eqnarray} 
assuming a constant electron density in the thick disk. In this work we take it to be $n_e=0.01~\text{cm}^{-3}$ \citep{yao_new_2017}. An analytical derivation of this expression can be found in \citet{pshirkov_mapping_2013}. This value is insensitive to the coherent magnetic field, which would change the RM value but not its standard deviation over small angular scales. This method of estimating $\theta$ using RM measurements is the one we ultimately use. Following an analysis of RM measurements in \citet{schnitzeler_latitude_2010}, we assume $\sigma_{RM,MW}(b)\approx\frac{7}{\sin(b)}~\text{rad/m}^{2}$ where $b$ is the galactic latitude. We find \begin{equation}
\label{eq:theta_b}
    \theta(b)\approx 8.6^{\circ}\left(\frac{R}{3~\text{EV}}\right)^{-1}\frac{1}{\sin(b)},
\end{equation} consistent with Eq.~(\ref{eq:basic_random_deflection}). This expression becomes inaccurate near the Galactic plane, overestimating the RM, so we mask the low latitudes ($|b|<10^\circ$) in the data whenever relevant. 

It is worth noting that in the presence of coherent deflections, the total path length of a cosmic ray inside the disk will not be an ideal straight line and the actual random deflection might be larger or smaller than the one implied by the RM measurements (in practice, it would be smaller, since the coherent field pushes the rays away from the Galactic plane and out of the disk). This is not a major effect, but we take it into account. For each GMF model $m=1,2\dots8$ and for each energy bin $i$ we backtrack rays from Earth out through the disk using CRPropa, get their corrected path length and correct the deflection angle accordingly to get a deflection map $\theta_{m,i}(l,b)$. The final value depends on the height of the thick disk. We checked two values, $1~\text{kpc}$ and $2~\text{kpc}$, and found that our final results are not sensitive to this choice. In the rest of this work it is assumed to be $1~\text{kpc}$.

The rigidity distribution we obtain for nuclei (shown in Fig.~\ref{fig:rigidity}), for rays above $E_o=2\times10^{19}~\text{eV}$ has 90\% of the rays with $R>3.1~\text{EV}$, which corresponds to $\theta_{random}\approx16.7^\circ$ at latitude $b=30^\circ$. This value of $\theta_{random}$ is an \textit{upper limit}; naturally the deflections will be smaller at higher energies and even smaller for a proton UHECR model. Anisotropies on bigger scales would not be smeared away by random deflections and should be visible in the data. The LSS does show anisotropies on larger scales, and this is what ultimately lets us differentiate between heavy and light composition scenarios.

\subsection{EGMF}
Eq.~(\ref{eq:magdef}) applies to deflections by the EGMF as well. The EGMF is expected to be mainly concentrated in LSS filaments that fill a small fraction of the volume, while most of the volume is filled by "voids" of very weak field. The RMS of the deflection angle can be estimated to be \citep{kashti_searching_2008}
\begin{equation}
    \theta\approx5.9^\circ\left(\frac{f_V}{0.05}\frac{D}{100~\text{Mpc}}\frac{l}{100~\text{kpc}}\right)^{1/2}\frac{B}{1~\text{nG}}\left(\frac{R}{3~\text{EV}}\right)^{-1},
\end{equation}
where $f_V$ is the fraction of volume filled by $1$~nG filaments and $l$ is the field's correlation length. A 5~nG field was adopted in \citet{kashti_searching_2008} based on equipartition arguments, consistent with observations and LSS MHD simulations \citep{2003Klypin-LocalBSim,2005Dolag-LocalBSim,2008Ryu-XGB-Sim,2008KoteraLemoine-Bdeflect}. More recent analyses \citep{vazza_simulations_2017,aramburo-garcia_magnetization_2021} yield somewhat lower values, with $f_V=5\%$ for $B=1$~nG (note that $l=100$~kpc is an overestimate, since eddies of this size complete one revolution in Hubble time at $100\,{\rm km/s}$).

The much larger deflections obtained in the "magnetic horizon" papers \citep{gonzalez_magnetic_2021,abdul_halim_impact_2024} require very strong magnetic fields in the "Local Supercluster," $\sim0.1\mu$G with $\sim1$~Mpc correlation length up to a distance of $\sim20$~Mpc \citep{abdul_halim_constraining_2024}. While such strong fields may be viable in highly over-dense filaments ($\delta\rho/\rho\gtrsim20$), they are not viable as permeating the entire nearby volume out to 20~Mpc distance, where the average over-density is small, $\delta\rho/\rho\simeq1.2$.

We conclude that the EGMF deflections are unimportant compared to the GMF deflections. 

\section{The LSS imprinted anisotropy}
\label{sec:method}

Here we describe the method we used to simulate cosmic ray arrival maps using the energy loss approximation outlined above, and the test statistics we derive from them. In the next section we present the actual results of the simulations and compare them to the data.

\subsection{LSS} \label{sec:lss}
The distribution of UHECR sources is taken to follow the structure of the nearby universe. Ideally we would have a 3D density map we could draw the sources from. \citet{bister_constraints_2024}, for example, use CosmicFlows-2 \citep{hoffman_quasi-linear_2018} for this purpose. We are instead going to create the density map based on the 2MRS galaxy catalog.

First, since the distance of nearby galaxies can't be reliably inferred from the catalog because their peculiar velocities are too large, for galaxies with \(v<2000~\text{km/s}\) we instead take redshift-independent distances from the HyperLEDA database \citet{makarov_hyperleda_2014}. For any other galaxy, the local group velocity is first subtracted from the catalog velocity to calculate the redshift. 

We follow the method outlined in \citet{crook_groups_2007} to fill in the Galactic plane where the catalog is lacking.

Of course, the catalog is flux limited, so just summing the number of catalog galaxies in each volume element will give us a skewed result. Instead, we estimate the luminosity selection function of the catalog, first apply k-correction to the apparent magnitude of each galaxy, then divide it by the luminosity selection function at its redshift, summing for all galaxies in a volume element. This process is inspired by the analysis in \citet{erdogdu_reconstructed_2006}. This is done in redshift bins of \(dz=0.001\) and angular HEALPix pixels created with healpy \citep{zonca_healpy_2019} with nside=32, above a distance of \(0.5~\text{Mpc}\) to exclude extremely bright sources in the local group, and below \(200~\text{Mpc}\). Above \(200~\text{Mpc}\) the distribution is taken to be isotropic (the validity of this approximation is checked by moving the limit around and demonstrating that the results don't change by much).

The result is still not ideal because we have artificially magnified the few far away galaxies we can see instead of spreading the missing far away luminosity more realistically. To fix this, after the luminosity correction, we smear each redshift bin shell with a gaussian beam with \(\sigma\) equal to the mean distance between visible galaxies at this shell. This should roughly preserve the structure that we can see but still spread the missing luminosity around in a more representative manner.

The end result is then normalized to calculate the distribution at any point in space
\begin{equation}
    w(x,z)=1+\delta(x,z),
\end{equation}
where $w$ is the normalized density and $\delta$ is the overdensity. The normalization is such that the mean density \(w\) is 1. Here and henceforth \(x\) refers to the angular coordinates ($l$ and $b$) together.

One final important point is our method for considering various source classes. The 2MRS is not a true trace of the matter density in the universe, since it is biased towards infrared-bright galaxies. Other types of potential sources (for example SBGs, AGNs) will also not exactly follow the matter distribution but will be biased in different ways. To first order, any source class will instead follow the biased distribution

\begin{equation}
    w=1+b\delta
\end{equation}
where \(b\) is some constant, and \(\delta\) is the "true" matter distribution overdensity. For the distribution traced by the 2MRS, the bias parameter is estimated to be something around \(b\approx1.15\) \citep{erdogdu_dipole_2006}. Star-forming galaxies, for example, will have a higher \(b\), since they only exist above some threshold density. To check the correlation of cosmic rays against some source class, instead of using a small catalog of sources with its own set of selection biases, our idea is to just skew the distribution we get from the much more complete 2MRS by choosing a different \(b\), trusting that the sources follow the LSS anyway. Since we don't have access to the "true" overdensity \(\delta\), only to the density field of the 2MRS \(w_{2MRS}=1+b_{2MRS}\delta\), we instead transform the field according to
\begin{equation}
    w=b_1(w_{2MRS}-1)+1
\end{equation}
by varying the parameter \(b_1=b/b_{2MRS}\).

\subsection{Generating arrival maps}
We vary the following parameters between realizations to determined their impact on the data:
\begin{itemize}
    \item \(s_0\), the source density in \(\text{count}/\text{Mpc}^3\). 
    \item \(a(x)\), the geometrical exposure map of the detector. We check four different values: the exposure for the available Auger Phase 1 data ("Auger exposure"), a recent combined TA+Auger exposure map with values quoted from \citet{matteo_2022_2023} ("combined exposure"), an idealized version of this map where the TA has equal exposure to Auger ("ideal combined exposure"), and a theoretical isotropic exposure map of a detector with \(2000~\text{km}^2\times5~\text{yr}\times4\pi~\text{sr}\) total exposure ("isotropic exposure").
    \item \(b_1\), the bias factor as described in \S~\ref{sec:lss}. We use \(b_1=1\) for a source distribution following the 2MRS, \(b_1=1.7\approx2/1.15\) as an example of a source distribution with a stronger mass bias - for example, elliptical galaxies - and \(b_1=0\) for a true isotropic source distribution.
    \item $m$, the coherent GMF model. We try the 8 different models presented in \citet{unger_coherent_2024} as well as a scenario with no GMF.
    \item The source spectrum: protons with a power law injection spectrum vs. nuclei with a power law + cutoff spectrum (detailed in \S~\ref{sec:analytic_treatment}, Eq.~(\ref{eq:protonspectrum}) and Eq.~(\ref{eq:theirspectrum}) respectively).
\end{itemize}
For each choice of parameters we generate 10000 arrival maps, then calculate our test statistics from them.

To generate an arrival map, first sources are sampled from the source distribution (to simulate the cosmic variance in source positions). In each volume element of the local universe, the expected source count is
\begin{equation}
    \bar S=w(x,z)s_0dV
\end{equation}
where \(dV=drd\Omega d_A^2\) is the volume, \(s_0\) is the source density and \(w(x,z)\) is the bias inferred from the LSS as described in the previous section.

The number of sources in each volume element is drawn from a Poisson distribution with mean \(\bar S\). Then, for each resulting source count \(S\), the expected flux from this volume element is
\begin{equation}
    F(x,z)= \frac{\psi(z;E_o)(1+z)}{s_04\pi d_L^2}S,
\end{equation} where \(\psi\) is defined in Eq.~(\ref{eq:psi_dndr}). \(\psi\) depends on the radial distance, the energy bin, and the source model (protons vs. nuclei). The total flux $F(x)$ from the direction $x$ is this value, summed along the radial direction.

Next we include the effect of magnetic deflections. First, the coherent field induces some transformation on the directions, $F_{deflected}(x)=F(L_{m,i}(x))$. To simulate the random field's impact, for each direction $x$, we draw a randomly deflected direction $x^\prime$ from a Gaussian beam around $x$ with standard deviation $\theta_{m,i}(x)$. $\theta_{m,i}$ and $L_{m,i}$ were both described in \S~\ref{sec:magnet}. The resulting flux on Earth is given by
\begin{equation}
    F_{E}(x)=\sum_yF_{deflected}(y)~\text{where $y^\prime=x$}.
\end{equation}

Finally, the expected ray count from a direction $x$ is
\begin{equation}
    \bar N=a(x)F_E(x),
\end{equation}
where \(a(x)\) is the total exposure (area~\(\times\)~time) in each direction.

We draw the ray count from a Poisson distribution with mean \(\bar N\) from each volume element for each energy bin, and sum the counts along the radial direction to get the arrival map.

Now briefly neglecting magnetic deflections, the expected value of this double Poisson process is, unsurprisingly
\begin{align} \label{eq:mean_count}
    E[N]=&\frac{a(x)\psi(z;E_o)(1+z)\bar S}{s_04\pi d_L^2}\\=&a(x)\psi(z;E_o)w(x,z)(1+z)^{-3}\frac{d\Omega}{4\pi}dr,
\end{align}
independent of the source density, but the variance is
\begin{equation}
    V[N]=E[N](1+\frac{a(x)\psi(x;E_o)(1+z)}{s_04\pi d_L^2}),
\end{equation}
which is bigger than the variance \(V[N]=E[N]\) for a regular Poisson process. The difference, which inversely depends on \(s_0\), is exactly the contribution from the cosmic variance, which converges to zero for a large enough source density.

\subsection{The observables} \label{sec:tests}
For each random arrival map \(N(x)\) we calculate a few quantities, then present their distribution for any choice of parameters. The idea is that these quantities can be used as test statistics to compare the data against different models - most importantly against the proton and nuclei models, but also against different source distributions.

Below is an explanation of the test statistics and the motivation behind them.

\subsubsection{Correlation to large scale structure} \label{sec:testlarge}
\begin{figure*}
    \centering
    \includegraphics[width=1\linewidth]{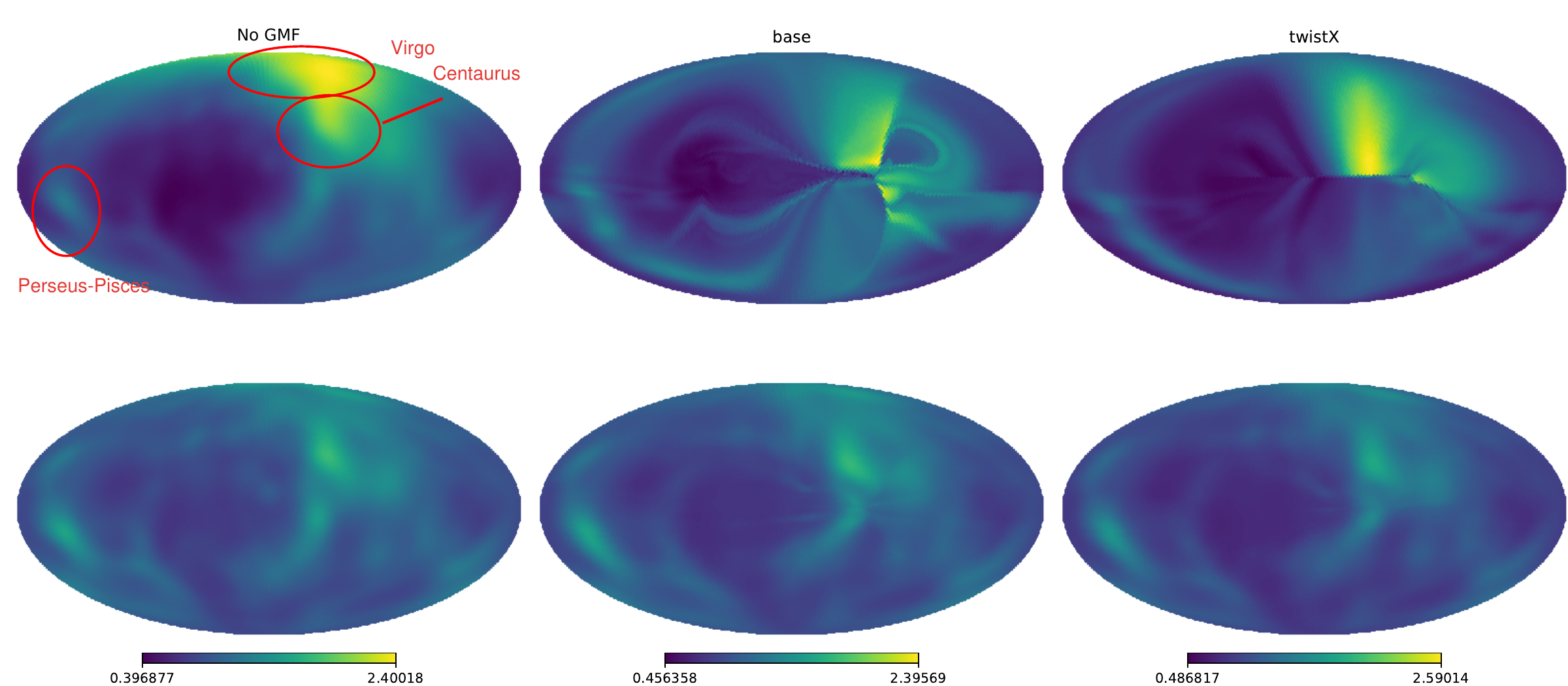}
    \caption{Mean arrival maps of heavy UHECRs (top row) and proton UHECRs (bottom row) with arrival energy $E_o>32~\text{EeV}$, without magnetic deflections (left) and with two of the eight UF23 models (middle and right), in Mollweide projection, galactic coordinates. Smearing by the random component of the GMF is not included. Some overdensities are identified with large nearby structures; the nearby Virgo cluster dominates the map of heavy UHECRs. After it, the brightest spots are the Centaurus supercluster and the Perseus-Pisces supercluster. Another notable feature is the very large void in the center-left of the map. The GMF smears the Perseus-Pisces signal along the galactic longitude, but generally preserves the feature intact. Rays originating in the Centaurus and Virgo clusters are deflected more significantly. The picture remains qualitatively similar for the other 6 GMF models.}
    \label{fig:meanmaps_iso}
\end{figure*}

\begin{figure*}
    \centering
    \includegraphics[width=1\linewidth]{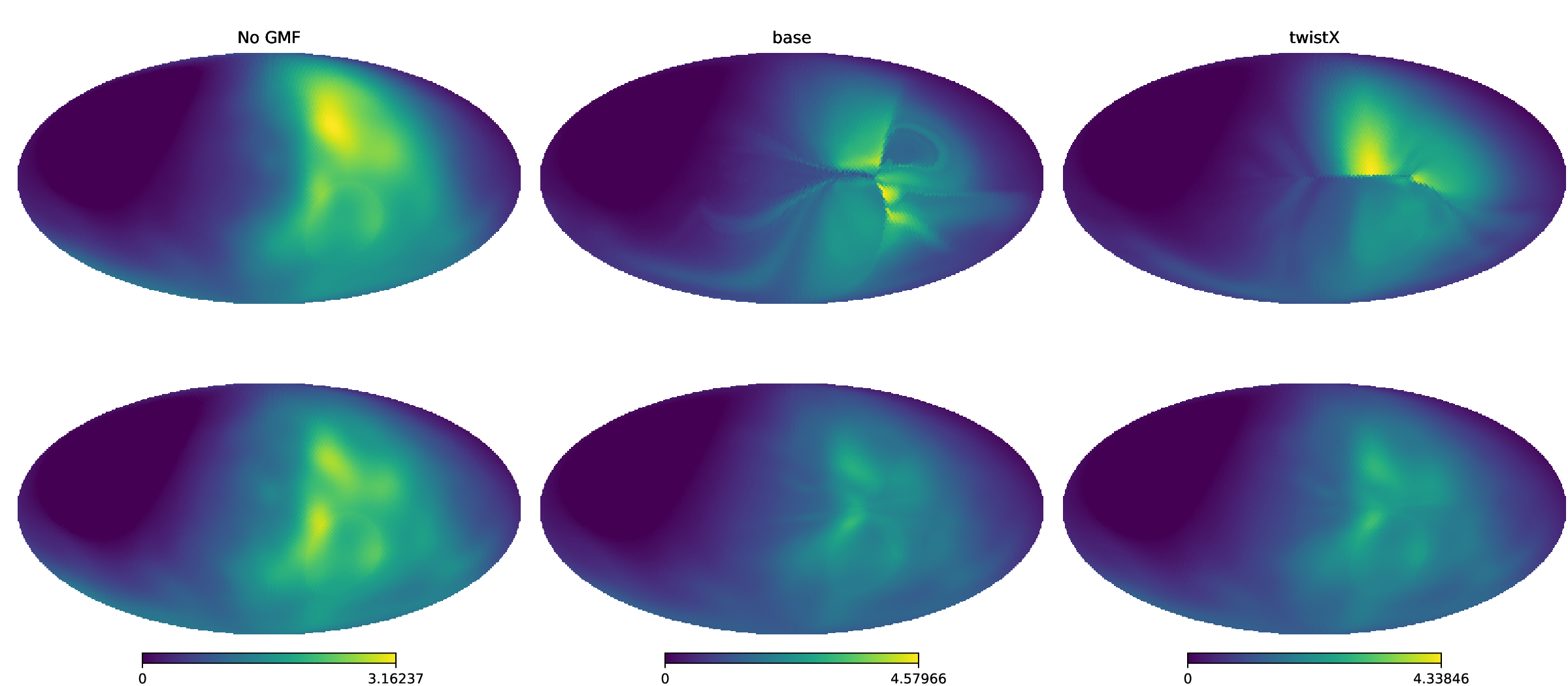}
    \caption{Same as Fig.~\ref{fig:meanmaps_iso}, except for an observer with the angular exposure of Auger.}
    \label{fig:meanmaps_auger}
\end{figure*}

For a given set of parameters \(p\) and energy bin we can calculate the \textit{mean arrival map} \(\Phi_p(x)\) by integrating Eq.~(\ref{eq:mean_count}) over redshift. \(\Phi_p\) is then normalized so its mean value per angular pixel is 1. These maps do include the coherent magnetic deflection but not the random magnetic smear. It is instructive to look at some of these maps to see what sorts of large scale structures we can expect to observe; see Fig.~\ref{fig:meanmaps_iso} and Fig.~\ref{fig:meanmaps_auger}.


Two important details are worth noting: first and most obvious, the anisotropies in the nuclei maps are much more notable, and exist on scales beyond $\theta_{random}=16.7^\circ$. The second is that even after the large coherent deflections are taken into account, it is still possible to point out over- and under- densities in some large regions of the sky.

The latter implies that a crude matched filter correlation on very large regions of the sky could still be a useful test that identifies the LSS signature and differentiates between nuclei and protons. We define five large angular bins that are described in Table~\ref{tab:bigbins}.

\begin{table}
    \centering
    \begin{tabularx}{\linewidth}{*4{>{\centering\arraybackslash}X}}
        \hline
        Name & $l$ & $b$ & dominant structure \\
        \hline
        $X_1$& $-$ & $45^\circ <b$ & Virgo\\
        \hline
        $X_2$& $100^\circ<l<180^\circ$ & $b<45^\circ$ & Perseus-Pisces\\
        \hline
        $X_3$& $0^\circ<l<100^\circ$ & $b<45^\circ$ & Void\\
        \hline
        $X_4$& $270^\circ<l<360^\circ$ & $b<45^\circ$ & Centaurus\\
        \hline
        $X_5$& $180^\circ<l<270^\circ$ & $b<45^\circ$ & Void\\
        \hline
    \end{tabularx}
    \caption{Large angular bins used for correlation with the LSS. The region $|b|<10^\circ$ is masked out from all of the bins.}
    \label{tab:bigbins}
\end{table}

 We divide the data to these bins, and calculate
\begin{equation} \label{eq:test_t}
    T_p=\sum_{X_i}N(X_i)\ln\Phi_p(X_i)/\sum_{X_i}N(X_i).
\end{equation}
This quantity represents how similar the map is to the expected map for a given model. The idea is that the choice of large bins will make this result independent of the specific, unknown deflections by the GMF, since the total count in each large bin will remain similar after the deflections.

The actually interesting quantity that we will consider is in fact \(T=T_p-T_q\) for two sets of parameters \(p\) and \(q\), which represents how much better an arrival map fits the model \(p\) than the model \(q\). After some trial and error, we find that choosing \(\Phi_p\) to be a map of nuclei, \(b_1=1\), and \(\Phi_q\) to a map of protons, \(b_1=1\) with no magnetic deflections for a given exposure map as a test statistic works well for testing both composition and source bias, where a higher \(T\) score correlates with higher anisotropy, or source "clumpiness".

Since the bins are so large, the random smearing is completely irrelevant here. On the other hand, these bins are not as useful when looking only at on hemisphere, for example when considering data from Auger alone. It is too easy for the coherent field to deflect nuclei across the Auger exposure. Values for $\Phi$ for several models are given in Table~\ref{tab:binphis}.

\begin{table*}
    \centering
    \begin{tabularx}{\linewidth}{*6{>{\centering\arraybackslash}X}}
        \hline
        Model & $\Phi_1$ & $\Phi_2$ & $\Phi_3$ & $\Phi_4$ & $\Phi_5$ \\
        \hline
        isotropic exposure, nuclei & $1.612$ & $0.856$ & $0.657$ & $1.166$ & $1.037$ \\
        \hline
        isotropic exposure, protons & $1.151$ & $1.001$ & $0.867$ & $1.090$ & $0.994$\\
        \hline
        ideal combined exposure, nuclei & $1.744$ & $0.968$ & $0.650$ & $1.066$ & $0.944$\\
        \hline
        ideal combined exposure, protons & $1.243$ & $1.122$ & $0.858$ & $1.004$ & $0.908$\\
        \hline
    \end{tabularx}
    \caption{Values of $\Phi$ (ray count inside a bin normalized to the area of the bin) for protons and nuclei, both for theoretical isotropic exposure and an ideal combined exposure (where the TA has equal exposure to Auger).}
    \label{tab:binphis}
\end{table*}

\subsubsection{Entropy} \label{sec:testentropy}
The large coherent deflections forced us to choose large angular bins in the method outlined above and as a result, a lot of information was lost. In principle, however, while the coherent deflections move the anisotropies around, the anisotropies are all still there in the map somewhere, aside from magnification/demagnfication effects. A very simple way of quantifying how uniform the direction distribution is is calculating the entropy of the arrival map,

\begin{equation}
    S=-\sum_{x_i}\tilde N(x_i)\log \tilde N(x_i).
\end{equation}
Here $\tilde N$ is the normalized, smeared arrival map. The normalization is coming from the usual definition of entropy over a distribution. "Smearing" here refers to smoothing the arrival with a Gaussian beam with standard deviation $\theta_{random}=16.7^\circ$. This step is important as it "washes out" any smaller scale anisotropies. Since the random deflections of protons are expected to be smaller than $\theta_{random}$, protons might actually look less isotropic on these scales; this smearing assures that we are only looking at anisotropies on scales where nuclei are supposed to dominate. We also mask the pixels with $|b|<10^\circ$ before doing this calculation to not be sensitive to the uncertainties near the Galactic plane.

Entropy is not necessarily an optimal test statistic, and we have found that it is less effective with a full sky exposure than in the Auger exposure alone because the map is overall more isotropic. For this reason, for the ideal combined exposure scenario, we calculate this "effective entropy" instead:
\begin{equation} \label{eq:effectiveentropy}
    S_E=\sum_{X_i} S(X_i)
\end{equation}
which is the sum of the entropies calculated separately for each of the large angular bins described in Table~\ref{tab:bigbins}. Since each of them alone exhibits more anisotropy than all of them do combined, this is a stronger metric of anisotropy. It is still not optimal, but as we will show, this quantity is sufficient for differentiating between protons and nuclei.

\subsubsection{Inter-energy correlation} \label{sec:testenergy}
For each realization, we calculate the arrival maps \(N_{32}(x),N_{42}(x)\) for rays in the energy range \(E>3.2\times10^{19}~\text{eV}\) and \(E>4.2\times10^{19}~\text{eV}\) respectively. Both maps are then smeared with a gaussian with \(\sigma=\theta_{random}\) and normalized to produce $\tilde N_{32}(x),\tilde N_{42}(x)$, then correlated to produce
\begin{equation} \label{eq:test_c24}
    C_{32,42}=\sum_{x_i}\tilde N_{32}^*(x_i)\tilde N_{42}^*(x_i)
\end{equation}
where the sum is for every angular pixel outside the Galactic plane ($|b|<10^\circ$). This quantity represents how much the map stays similar between these different energies.

We believe nuclei will show stronger correlation between different energies, since nuclei across the spectrum should come from similar distances and therefore similar sources. One could worry that the coherent deflection that differs between energy ranges will be enough to decorrelate the signals. If both energy ranges were deflected the same by the coherent field, then the actual configuration of the field does not matter and (bar magnification effects) the correlation of two full sky maps will remain the same, but of course different energy ranges are not deflected the same way. However, note that the expected difference between coherent deflection angles in different energies is

\begin{equation}
    \Delta\theta_{coherent}\approx\frac{Bd}{R_{32}}-\frac{Bd}{R_{42}}=\theta_{32}(1-\frac{R_{32}}{R_{42}}),
\end{equation}
where \(R\) are typical rigidities for these energy ranges. For protons, the rigidity ratio is exactly $\frac{32}{42}$, and typical values for the deflection angle outside the Galactic plane give us an overall \(\Delta\theta\approx\frac{1}{4}\theta_{32}\) that is smaller than \(\theta_{random}\). For nuclei, using our calculation for the rigidity distribution presented in Fig.~\ref{fig:rigidity}, by looking at the ratio of the means, the value seems to be closer to \(R_{32}/R_{42}\approx0.82\), giving \(\Delta\theta=0.18\theta_{32}\), so even with a larger inherent deflection, the weak dependence of the rigidity on energy means that for nuclei too the difference in deflection between the two energy ranges is comparable to the random smearing. If the coherent deflection is not in a straight line, the angular difference between the two energy ranges is even smaller.

\subsubsection{Dipole} \label{sec:diptest}
We use the method outlined in \citet{aublin_generalised_2005} to reconstruct the dipole direction $\vec d$ and amplitude $\alpha$ for each arrival map. $\vec d$ is a unit vector and $\alpha$ is a dimensionless number, the dipole amplitude relative to the monopole. The dipole is not expected to be sensitive at all to deflections by the random component of the GMF since these happen on much smaller angular scales, but both the direction and amplitude do depend on the details of the coherent component.

\section{Results} \label{sec:results}
\subsection{General comments} \label{sec:results_general}
While we do reproduce the UHECR spectrum somewhat accurately (see \S~\ref{sec:spectrum}), reproducing the exact number of observed cosmic rays is not easy as it depends on the exact positioning of the sources in the universe as well as accounting for magnification by the GMF. Ultimately both our nuclei and proton models do undershoot the "correct" ray count by some. This means that almost all of the histograms presented in this section are artificially wider than they should be, as an artifact of the (incorrectly) decreased ray count. This makes all the results slightly more conclusive than they appear in the plots. The p-values presented below do not account for this effect, which is quite small anyway (at most a factor of $1.1$). This correction only applies to the \textit{widths} of Gaussians, not their positions, since all the test statistics we use are normalized for ray count.

Additionally, the Auger data that we use suffer from $\approx14\%$
systematic uncertainty in energy calibration \citep{abreu_arrival_2022}. In all numerical results described below that use the available Auger data, we account for this uncertainty by comparing the data with simulations of rays at energies above and below the reported 32~EeV threshold, and report only the least-constraining p-values.

Finally, a comment is in place on the robustness of the results to the strength of the random GMF component; while the results are quite sensitive to the choice of $\theta_{random}=16.7$ \textit{used in the definition of the entropy test}, they are \textit{not} as sensitive to the actual values $\theta_{m,i}$ we use when generating arrival maps, both defined in \S~\ref{sec:magnet}. This is because $\theta_{random}$ is a conservative upper bound. For example, the results discussed in the entropy subsection below retain a $\sim90\%$ CL even after multiplying the deflection angle $\theta_{m,i}$ by 2.

\subsection{Correlation to large scale structure} \label{sec:results_corr}
\begin{figure*}
    \centering

    \includegraphics[width=1\linewidth]{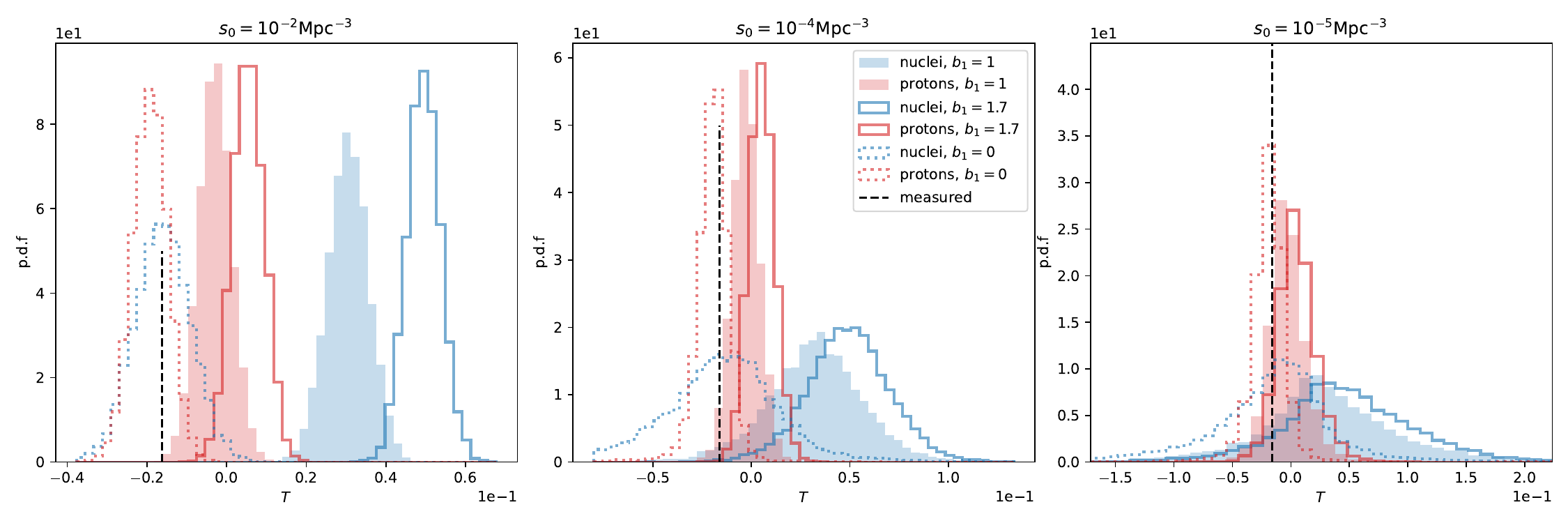}
    \caption{The LSS correlation statistics, $T$, for $E>32~\text{EeV}$ arrival maps, for an Auger-like exposure, different source densities, different source bias parameters, without coherent magnetic deflections. The single value calculated for the available data is marked with a black dashed line.}
    \label{fig:results_large_auger_simple}
    
    \bigskip

    \includegraphics[width=1\linewidth]{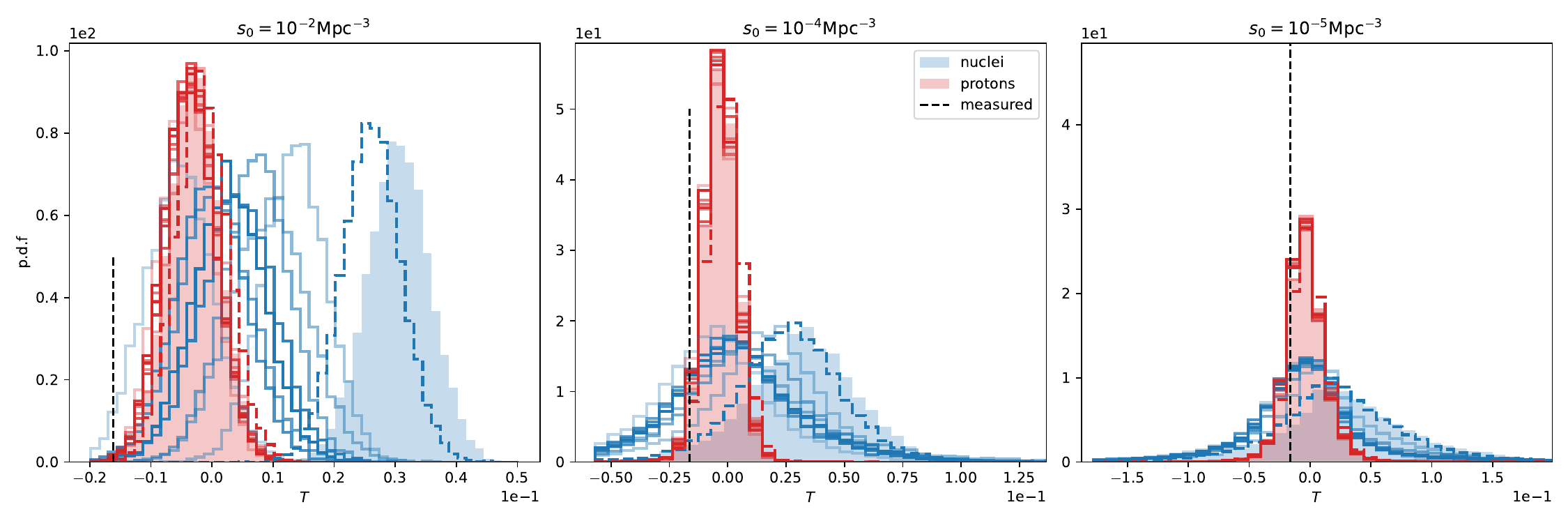}
    \caption{Same as Fig.~\ref{fig:results_large_auger_simple}, except restricting to $b_1=1$ and looking at various GMF models. The filled distributions are without coherent deflections, the outlined distributions are for the 8 different UF23 models. The dashed distribution is for the KST24 model.}
    \label{fig:results_large_auger_mag}

    \bigskip
    
    \includegraphics[width=1\linewidth]{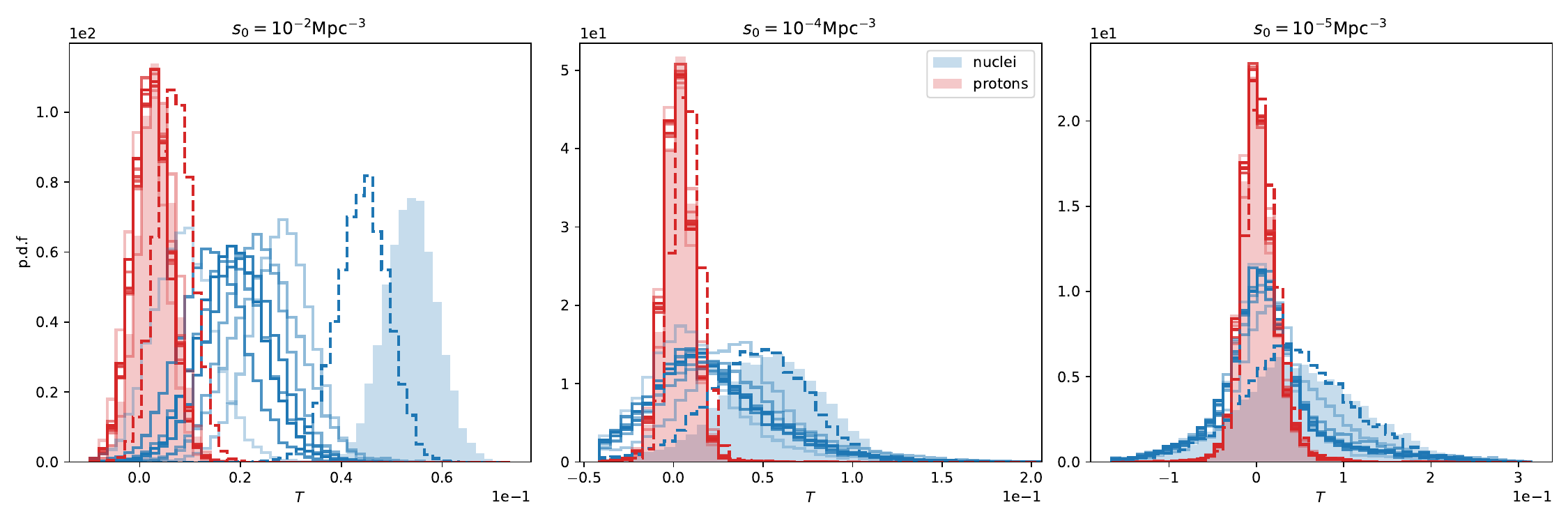}
    \caption{Same as Fig.~\ref{fig:results_large_auger_mag}, except for an ideal all-sky exposure - an idealized exposure map where the TA has equal integrated exposure to Auger.}
    \label{fig:results_large_ideal_mag}
\end{figure*}

Results for an Auger exposure are shown in Fig.~\ref{fig:results_large_auger_simple} and Fig.~\ref{fig:results_large_auger_mag}. The data are consistent with an isotropic source distribution. The non-detection of the LSS correlation signal implies that the source density must be fairly low. A source density of $s_0=10^{-2}~\text{Mpc}^{-3}$ is ruled out with a p-value of $<0.1\%$ even for the relatively isotropic proton model. The p-value increases beyond $10\%$ (when accounting for possible systematic error in the Auger energy calibration) for source densities $s_0\le 10^{-4}~\text{Mpc}^{-3}$. For heavy nuclei models, both low source density and GMF deflections are required to account for the observed $T$ value with $s_0\approx10^{-4}~\text{Mpc}^{-3}$, while $s_0\le 10^{-5}~\text{Mpc}^{-3}$ would be required in the absence of such deflections.

Coherent magnetic deflections erase the LSS signal from the nuclei arrival map, making the $T$ statistics inefficient in differentiating proton and heavy nuclei models. Fig.~\ref{fig:results_large_ideal_mag} shows the results for the ideal combined exposure. The composition signal is somewhat stronger, but not enough to allow drawing conclusions with a certainty of $2\sigma$ even in the best case of $s_0=10^{-2}~\text{Mpc}^{-3}$. At lower densities, the heavy-composition signal remains undetectable.

These results, specifically the fact that the coherent deflections erased the signal, show that ultimately this choice of large bins to try and "evade" the GMF impact did not prove useful (except for the KST24 GMF model). While it is possible that a smarter choice of bins would succeed where ours failed, we leave this to future work.

\subsection{Entropy} \label{sec:results_entropy}
\begin{figure*}
    \centering
    \includegraphics[width=1\linewidth]{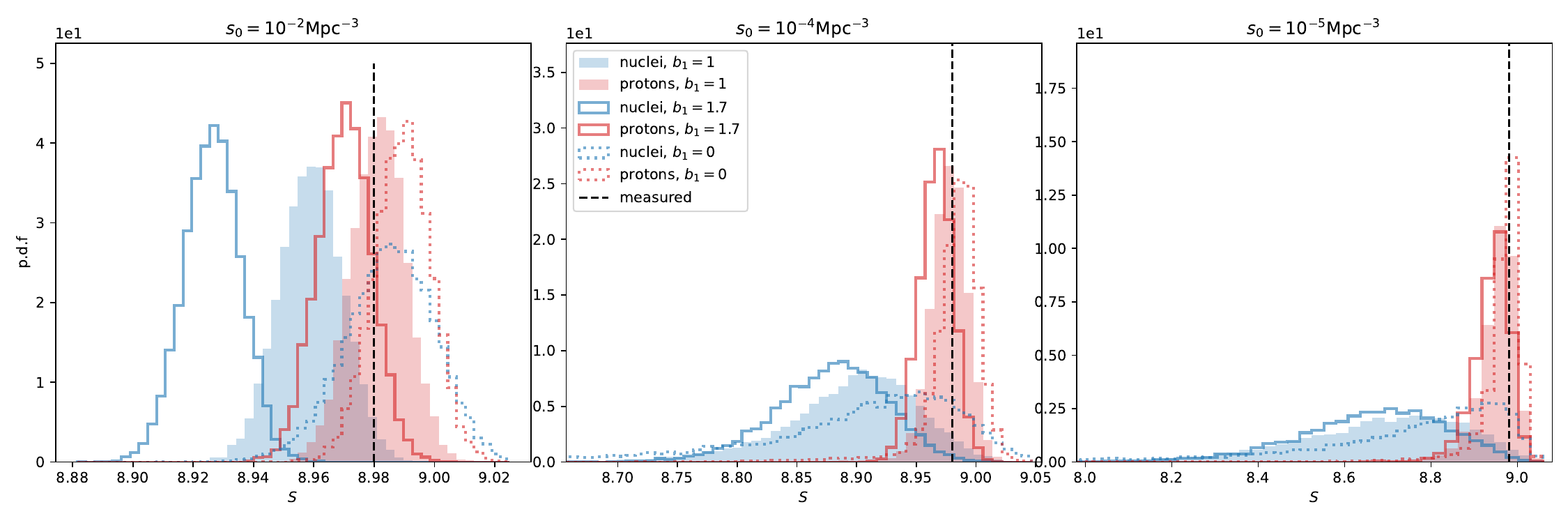}
    \caption{The entropy $S$ for $E>32~\text{EeV}$ arrival maps, for an Auger-like exposure, different source densities, different source bias parameters, without coherent magnetic deflections. The single value calculated for the available data is marked with a black dashed line.}
    \label{fig:results_entropy_auger_simple}

    \bigskip
    
    \includegraphics[width=1\linewidth]{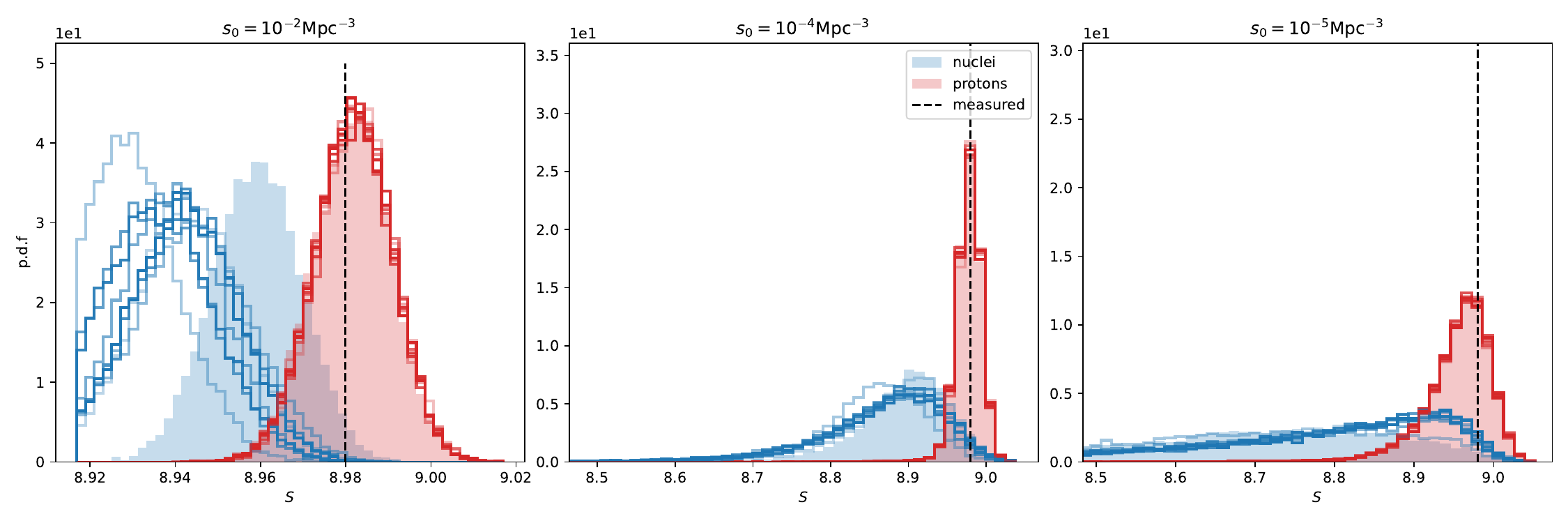}
    \caption{Same as Fig.~\ref{fig:results_entropy_auger_simple}, except restricting to $b_1=1$ and looking at various GMF models. The filled distributions are without coherent deflections, the outlined distributions are for the 8 different UF23 models.}
    \label{fig:results_entropy_auger_mag}

    \bigskip
    
    \includegraphics[width=1\linewidth]{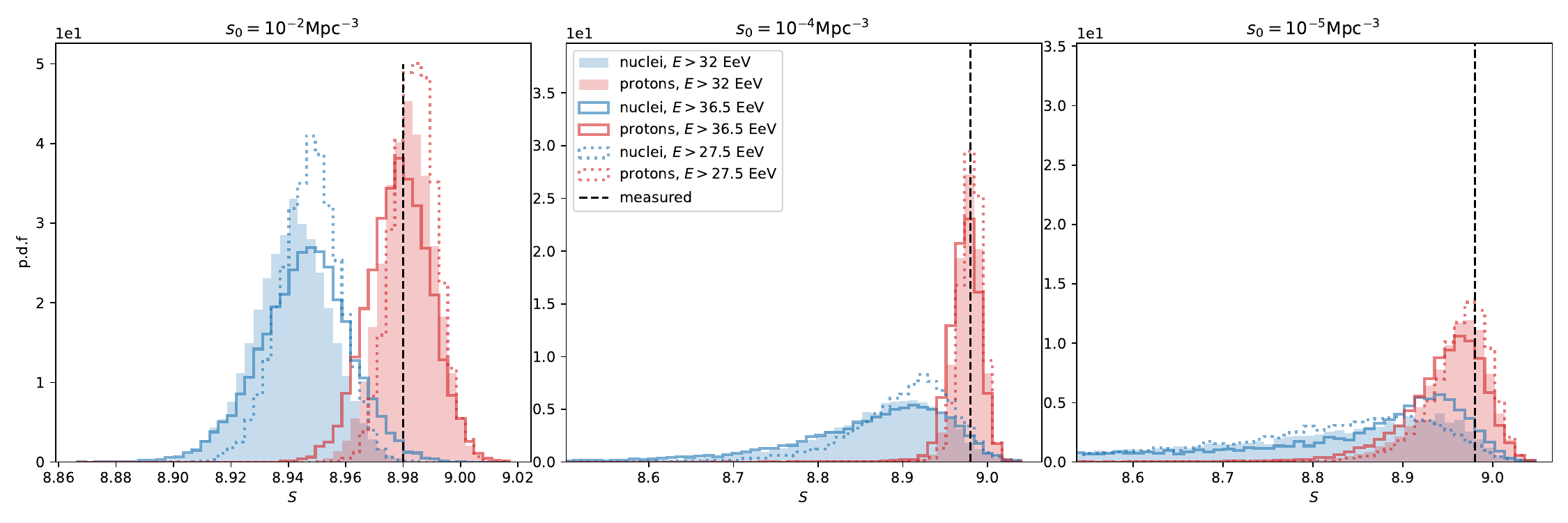}
    \caption{Same as Fig.~\ref{fig:results_entropy_auger_simple}, except restricting to $b_1=1$, choosing the UF23 base model and varying the energy threshold to simulate the effect of systematics in Auger energy measurements ($\sim14\%$).}
    \label{fig:results_entropy_auger_calib}
\end{figure*}

Results for an Auger exposure are shown in Fig.~\ref{fig:results_entropy_auger_simple}, Fig.~\ref{fig:results_entropy_auger_mag} and Fig.~\ref{fig:results_entropy_auger_calib}. As expected, nuclei models are more anisotropic and hence have less entropy. The GMF does not effect the measurement for protons at all, which is to be expected from looking at Fig.~\ref{fig:meanmaps_iso}, as their magnification is very weak. For nuclei, the GMF actually decreases the entropy, but it is important to note that this is a property of the arbitrary area in the sky that Auger can see; when taking the entire sky, the GMF actually increases the overall entropy.

Heavy nuclei models are ruled out for a source density of $s_0=10^{-2}~\text{Mpc}^{-3}$, and all eight UF23 models, with a p-value of $0.3\%$ or better. For a very low source density, $s_0=10^{-5}~\text{Mpc}^{-3}$, this number goes up to $4\%$ ($5\%$ when considering the possible systematic error in the Auger energy calibration).

\begin{figure*}
    \centering
    \includegraphics[width=1\linewidth]{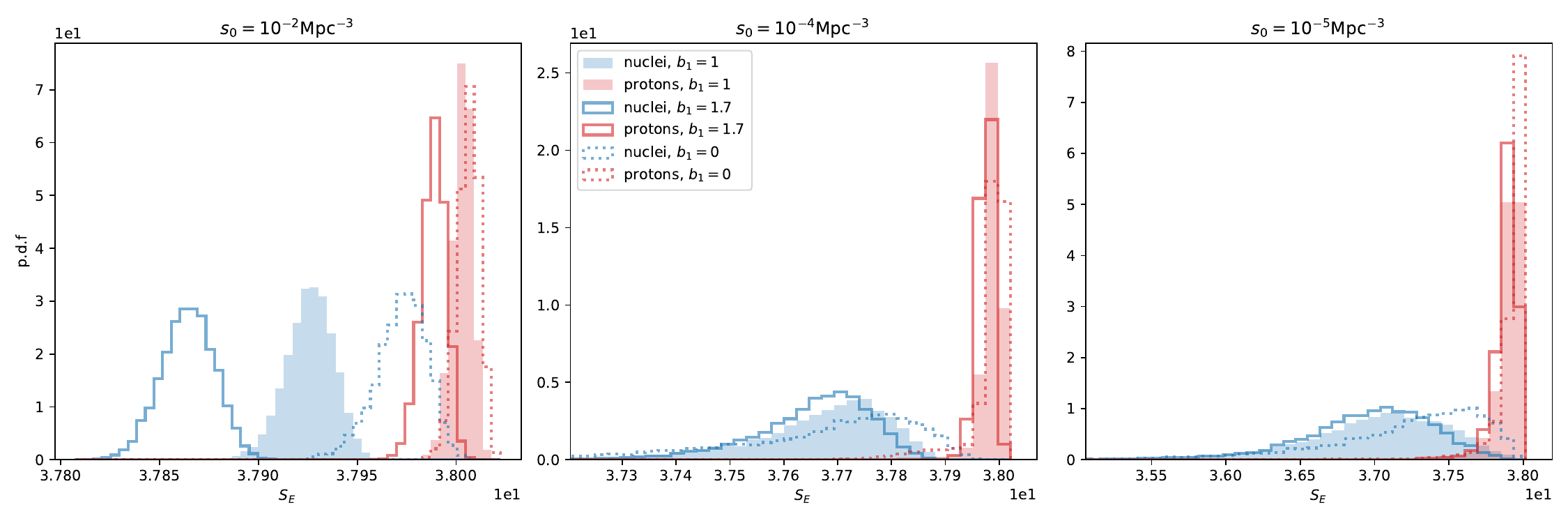}
    \caption{Same as Fig.~\ref{fig:results_entropy_auger_simple}, for the effective entropy (Eq.~(\ref{eq:effectiveentropy})) for an ideal all-sky exposure.}
    \label{fig:results_entropy_ideal_simple}

    \bigskip

    \includegraphics[width=1\linewidth]{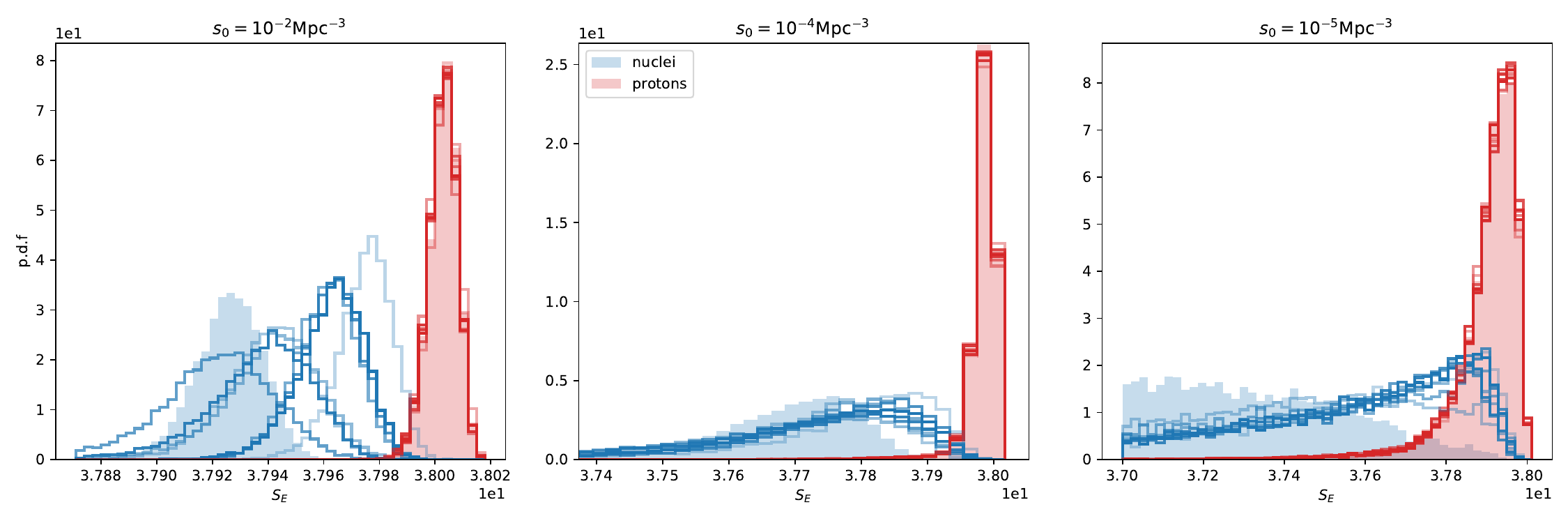}
    \caption{Same as Fig.~\ref{fig:results_entropy_auger_mag}, for the effective entropy (Eq.~(\ref{eq:effectiveentropy})) for an ideal all-sky exposure.}
    \label{fig:results_entropy_ideal_mag}
\end{figure*}

Results for an ideal all-sky exposure are shown in Fig.~\ref{fig:results_entropy_ideal_simple} and Fig.~\ref{fig:results_entropy_ideal_mag}, using the effective entropy described in Eq.~(\ref{eq:effectiveentropy}). Data from the entire sky would not improve the results much, as the strongest anisotropies are already visible in the Auger data. Demagnification by the GMF is more pronounced in the northern hemisphere; hence, the GMF ends up increasing entropy. Still, the slight improvement might be enough to start constraining $b_1$. This strange behavior, where the increased exposure does not actually increase the signal (and seemingly lowers it in the high-density, GMF-included case) proves what we already knew, that this entropy measure is not an optimal test statistic. It's important to note that this does not imply using it as a test statistic is incorrect, just that in theory there exists a better discriminator between the models that does not suffer this problem.

\subsection{Inter-energy correlation}\label{sec:results_energy}
\begin{figure*}
    \centering
    \includegraphics[width=1\linewidth]{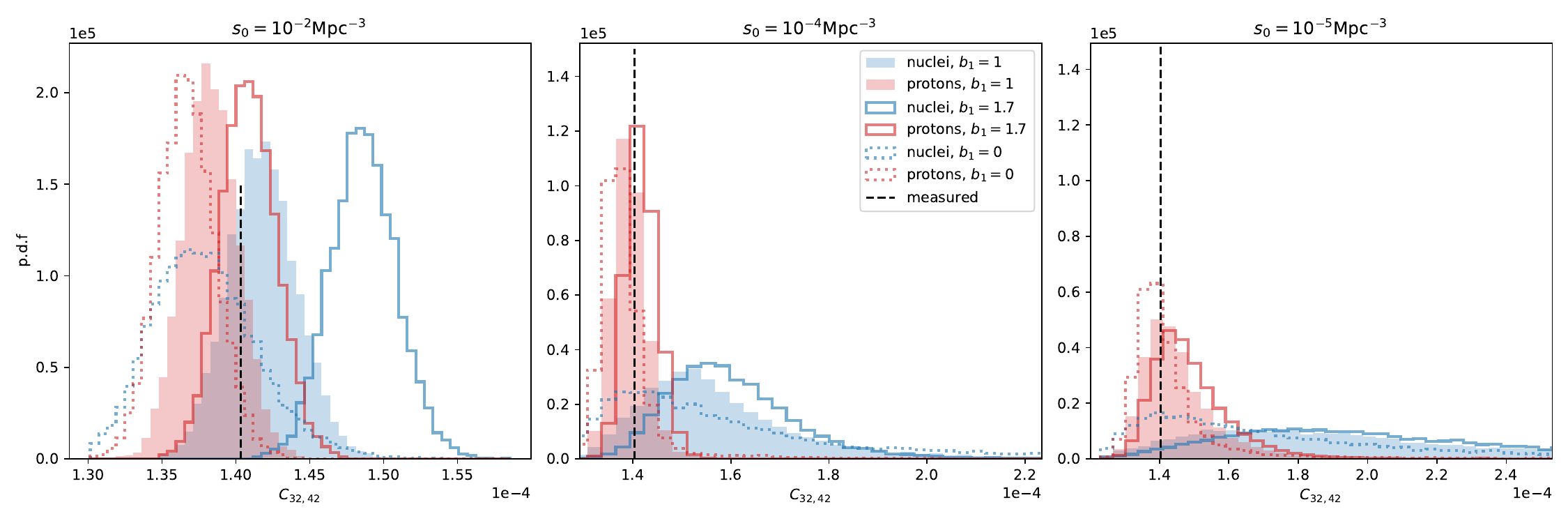}
    \caption{The correlation $C$ between $E>32~\text{EeV}$ and $E>42~\text{EeV}$ arrival maps, for an Auger-like exposure, different source densities, different source bias parameters, without coherent magnetic deflections. The single value calculated for the available data is marked with a black dashed line.}
    \label{fig:results_ecorr_auger_simple}

    \bigskip
    
    \includegraphics[width=1\linewidth]{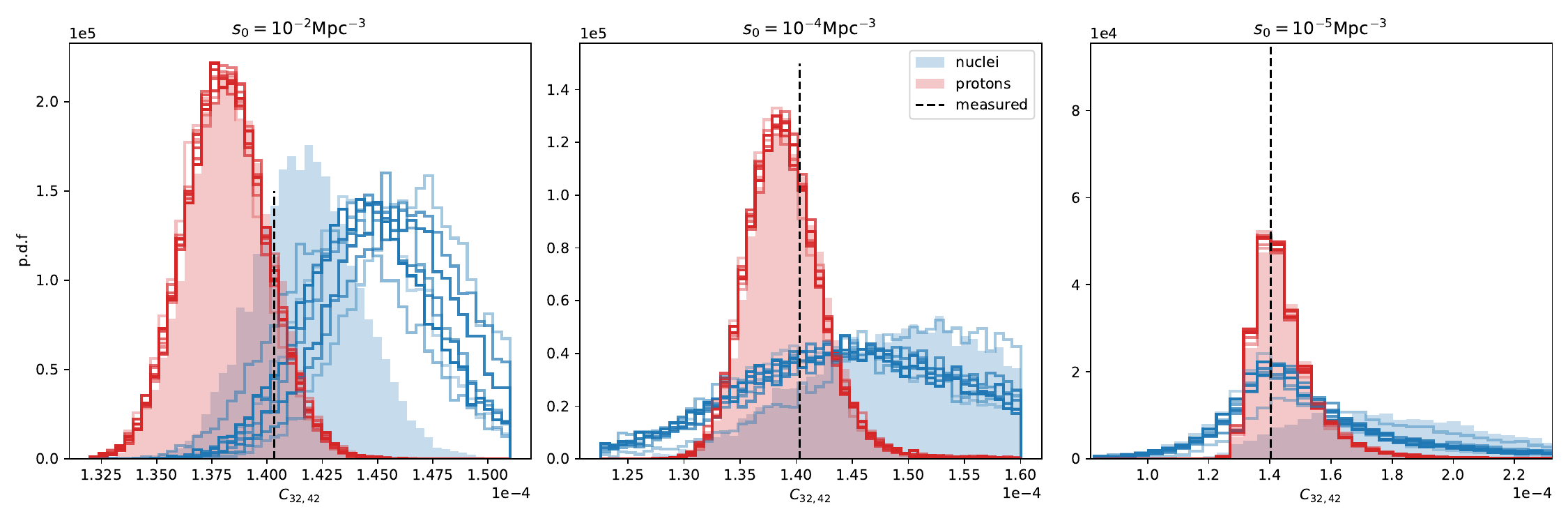}
    \caption{Same as Fig.~\ref{fig:results_ecorr_auger_simple}, except restricting to $b_1=1$ and looking at various GMF models. The filled distributions are without coherent deflections, the outlined distributions are for the 8 different UF23 models.}
    \label{fig:results_ecorr_auger_mag}

    \bigskip
        
    \includegraphics[width=1\linewidth]{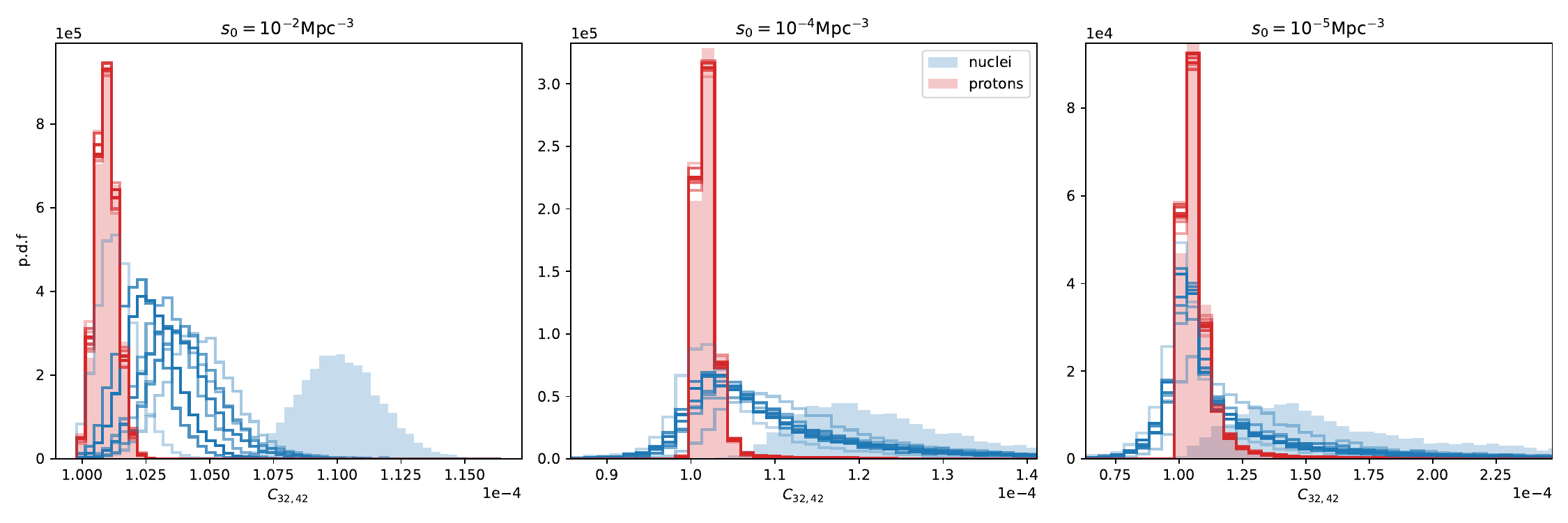}
    \caption{Same as Fig.~\ref{fig:results_ecorr_auger_mag}, except for an ideal all-sky exposure.}
    \label{fig:results_ecorr_ideal_mag}
\end{figure*}

Results for an Auger exposure are shown in Fig.~\ref{fig:results_ecorr_auger_simple} and Fig.~\ref{fig:results_ecorr_auger_mag}. While in principle there is some signal ($C$ is bigger for nuclei than for protons), it is not very significant in the data that we have ($\sim10\%$ percent p-value for the highest source density).

Fig.~\ref{fig:results_ecorr_ideal_mag} shows results for the ideal combined exposure. The coherent deflections keep the signal just as weak. It is worth noting that while the GMF magnifies the signal for the Auger half of the sky, it actually weakens it for all-sky exposure (a similar effect occurs with entropy, but less severely). This is partly because the Virgo cluster in the northern hemisphere, one of the closest and largest local structures, is severely affected by the GMF, and therefore, there is much less structure in the TA part of the sky.

\subsection{Dipole}\label{sec:results_dipole}
\begin{figure*}
    \centering
    \includegraphics[width=1\linewidth]{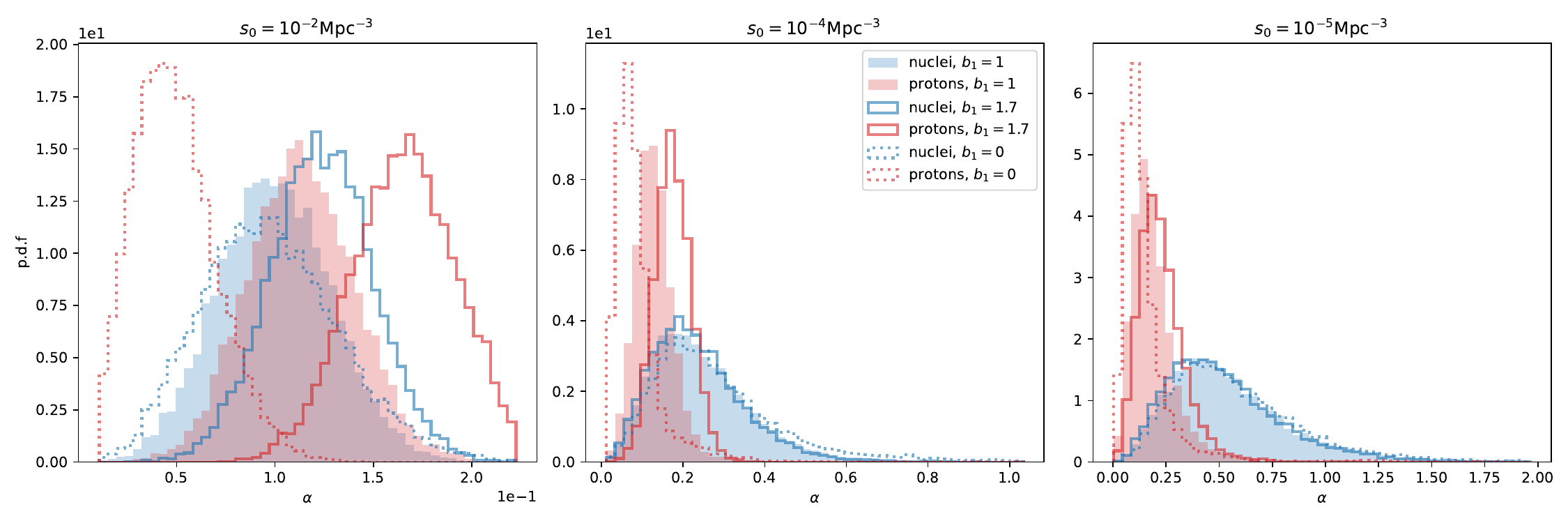}
    \caption{The dipole amplitude $\alpha$ for $42~\text{EeV}>E>20~\text{EeV}$ arrival maps, for a combined Auger+TA exposure, different source densities, different source bias parameters, with the base GMF model. Results look qualitatively similar for the other models.}
    \label{fig:results_dip_combined_simple}

    \bigskip

    \includegraphics[width=1\linewidth]{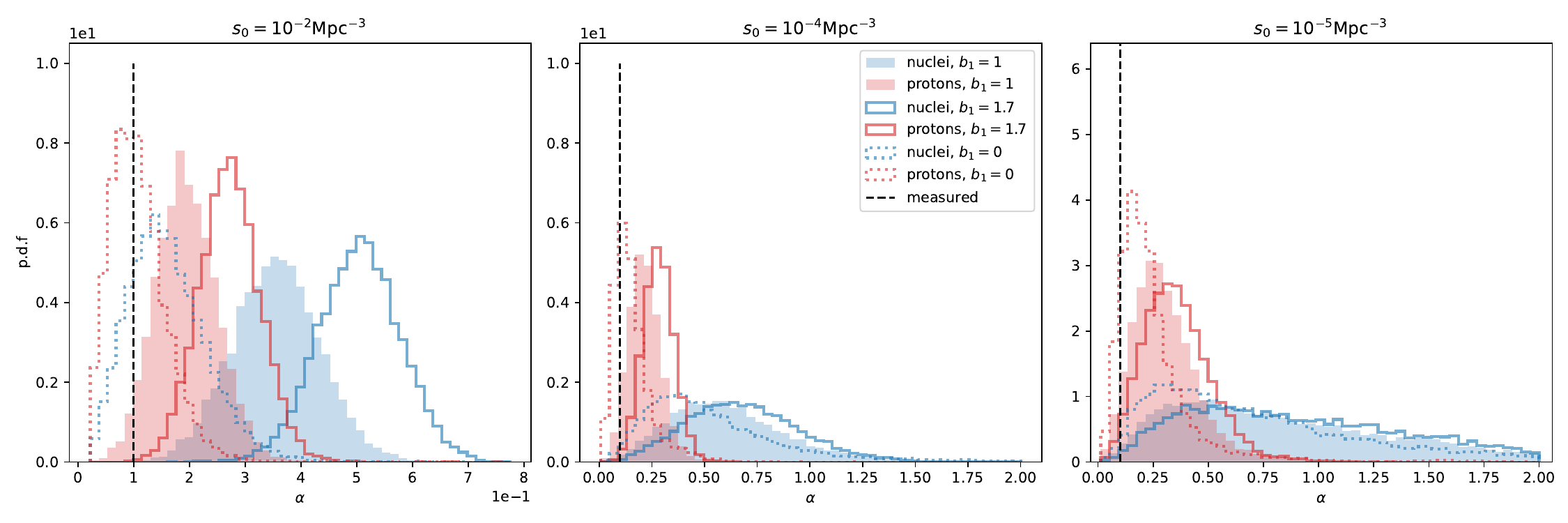}
    \caption{The dipole amplitude $\alpha$ for $E>32~\text{EeV}$ arrival maps, for an Auger-like exposure, different source densities, different source bias parameters, with the base GMF model. The single value calculated for the available data is marked with a black dashed line.}
    \label{fig:results_dip_auger_simple}

    \bigskip

    \includegraphics[width=1\linewidth]{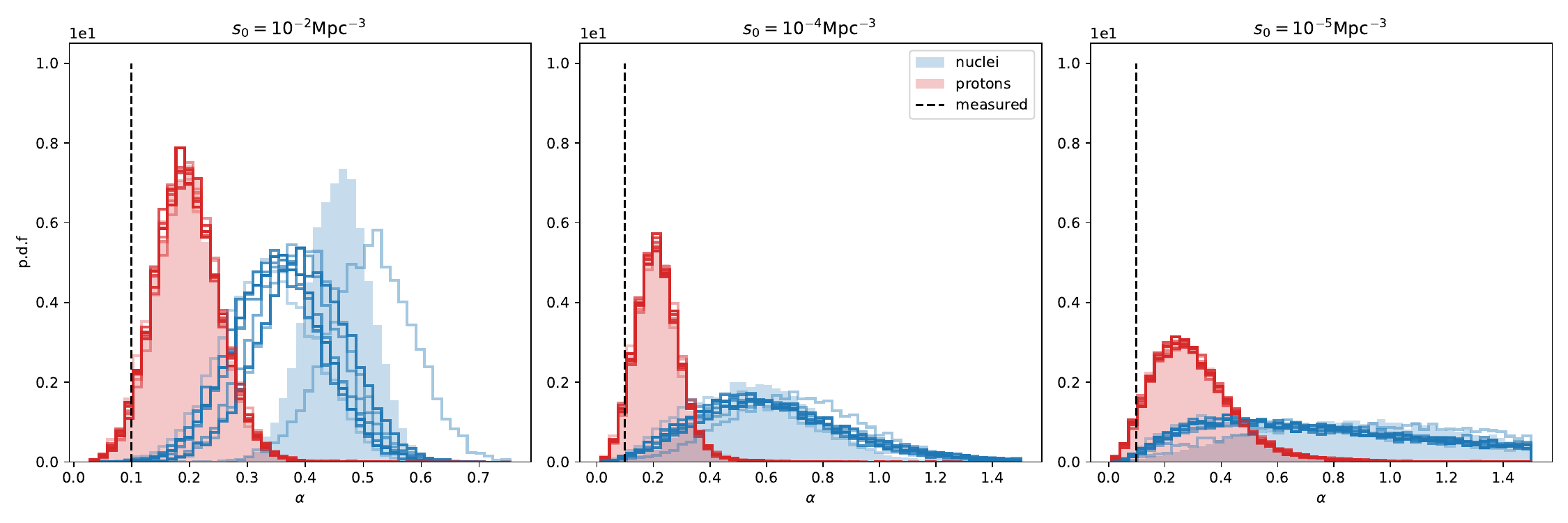}
    \caption{Same as Fig.~\ref{fig:results_dip_auger_simple}, except restricting to $b_1=1$ and looking at various GMF models. The filled distributions are without coherent deflections, the outlined distributions are for the 8 different UF23 models.}
    \label{fig:results_dip_auger_mag}
\end{figure*}

Results for a combined Auger and TA exposure are shown in Fig.~\ref{fig:results_dip_combined_simple}. The energy range is similar to that analyzed in \citet{urena_new_2025}, and the value they calculate for the data, $d\approx0.1$, is consistent with our simulation results. Their exact value is sensitive to the choice of energy calibration between TA and Auger, but the relatively wide distribution we observe suggests it would still remain consistent with a different calibration choice. It is apparent that the dipole amplitude is the same in the light and heavy UHECR models and therefore cannot be used to distinguish between them, nor does it allow us to constrain the source density.

Results for the public Auger data, at $E>32$~EeV, are shown in Fig.~\ref{fig:results_dip_auger_simple} and Fig.~\ref{fig:results_dip_auger_mag}. There is a tension between the observations, which are consistent with isotropy, and the model's predictions: the predicted values are larger than the observed value even for protons, which are much more isotropic than nuclei. Proton models are inconsistent with the data at 94\% (92\%) CL for a source density of $10^{-2}~\text{Mpc}^{-3}$ ($10^{-4}~\text{Mpc}^{-3}$), while heavy-nuclei models are inconsistent with the data at $>99\%$ ($\approx$99\%), with the exact value depending on the GMF model used. These values account for possible systematic errors in the Auger energy calibration.

\begin{figure*}
    \centering
    \includegraphics[width=1\linewidth]{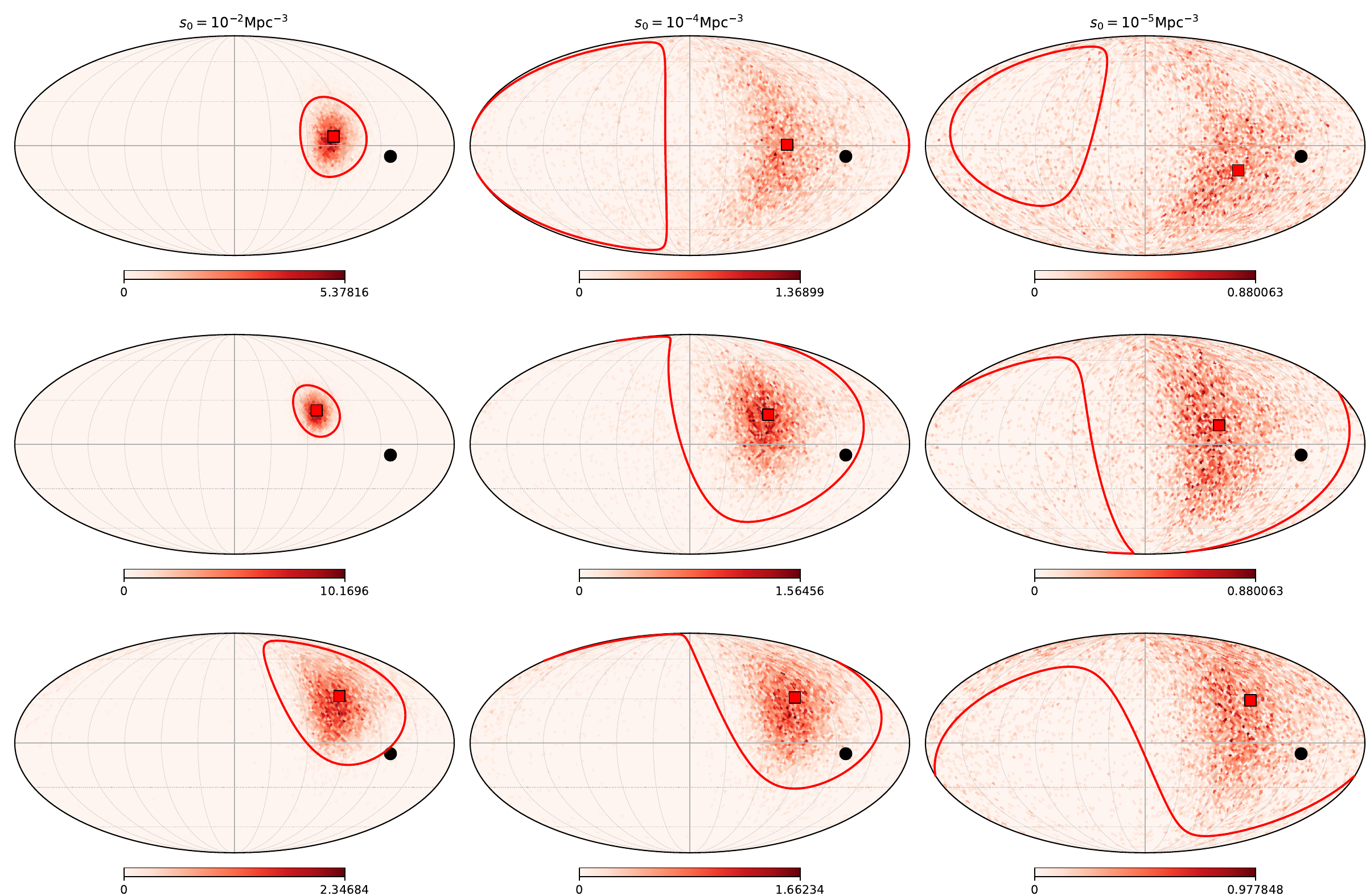}
    \caption{
    The distribution of the reconstructed dipole direction for $E>32~\text{EeV}$ arrival maps in galactic coordinates, for an Auger-like exposure, with different source densities. The top row is for a heavy composition and the base GMF model, the middle row is for heavy composition and the twistX GMF model, the bottom row is for a proton composition and the base GMF model. Units are probability per square radian. The proton distribution is not sensitive to the choice of the GMF model, while the heavy distribution is. The red contour is the circle around the barycenter of the distribution that contains 90\% of the points, the black point is the reconstructed dipole direction of the public data.}
    \label{fig:results_dip_directions}
\end{figure*}

\begin{figure}
    \centering
    \includegraphics[width=1\linewidth]{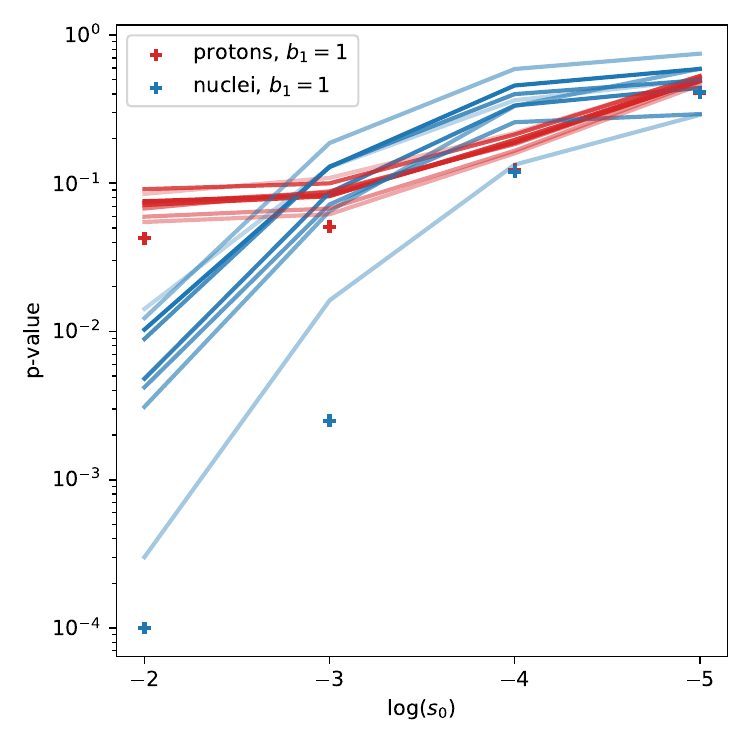}
    \caption{The probability of observing a reconstructed dipole direction farther from the predicted mean direction than the observed one, for different models. The crosses are in the case of no coherent GMF deflections, the lines are for the 8 UF23 GMF models. }
    \label{fig:results_dip_pval}
\end{figure}

A few illustrative distributions of the dipole direction are shown in Fig.~\ref{fig:results_dip_directions}. Fig.~\ref{fig:results_dip_pval} shows how the "p-value", in this case defined to be the fraction of the distribution farther from the barycenter than the observed direction, depends on the source density. It is apparent that low source densities are not constrained at all while a higher source density ($10^{-2}~\text{Mpc}^{-3}$) is ruled out with $\sim93\%$ CL for a light composition and $\sim99\%$ CL for a heavy composition - however, in the case of a heavy composition, the predicted direction significantly depends on the choice of GMF model, so it is difficult to draw conclusions with a high certainty.

\section{Comparison to previous works} \label{sec:comparison}
Below, we address some earlier works that considers anisotropies in the UHECR arrival map.

In contrast with our findings, a paper by the TA collaboration \citep{telescope_array_collaboration_mass_2024} finds that a heavy composition is favored by the arrival direction distribution observed by the TA at energies $>10~\text{EeV}$ on the grounds that a strong magnetic deflection is required to explain the isotropy of the data (a light composition would still be possible with an unreasonably strong EGMF). First, we note that in our analysis we use the public Auger data, whereas applying our methods to the unpublished TA data may yield different results (we also note that the TA paper makes statements about rays with $E>100$~EeV, while we find that, at these energies, the statistics are insufficient for our tests to yield significant constraints). Moreover, we analyze an aspect of isotropy (entropy) that differs from those analyzed in the TA paper; thus, the implications of the data being "more isotropic" may also differ. 

Nevertheless, it should be noted that there is some uncertainty in the statistical significance of the evidence presented for heavy composition. The analysis is based on calculating a "test statistic" (TS) for the data that depends on a deflection parameter $\theta_{100}$ (representing the typcial deflection at 100~EeV), TS$(\theta_{100})$, determining the value of $\theta_{100}$ minimizing TS$(\theta_{100})$, which we denote $\theta^D_m$, comparing it to the average value of this angle, $\theta_m^M$, obtained applying a similar procedure to simulated data produced by models with different composition and a given (fixed) GMF, and estimating the significance of a deviation between $\theta_m^M$ and $\theta_m^D$ by determining an uncertainty in $\theta^D_m$ from the data's TS$(\theta_{100})$ curve. It is not proven \citep[also in][which is the basis for the analysis]{kuznetsov_uhecr_2021} that this is a valid method of estimating the significance, i.e. that it is equivalent to comparing the deviation between $\theta_m^M$ and $\theta_m^D$ to the spread in $\theta_m^M$ values obtained in model realizations, which should include the impacts of a finite source number and GMF uncertainties (the inclusion of the former effect is attempted in their appendix, but the method of randomly selecting sources while applying the same TS$(\theta_{100})$ as before does not necessarily represent the actual variance in a finite source number model). In particular, for $E>100$~EeV the data's TS$(\theta_{100})$ has no minimum, leading to the claim that $\theta_m^M$ should exceed 200~deg thus excluding protons as reasonable candidates. Here it should be noted that such large deflections are also challenging for Fe, see eqs.~(\ref{eq:basic_random_deflection}, \ref{eq:theta_b}), and that in order to determine the significance of excluding protons in this method, the probability should have been calculated (by examining many model realizations) of obtaining a TS$(\theta_{100})$ curve with no minimum for 19 events (such a calculation was not presented).

\citet{higuchi_insights_2026} examine rays with energy $>100~\text{EeV}$ in the Auger data, and argue that the apparent isotropy - specifically, the remarkably low number of observed cosmic ray pairs with angular separation $<3^\circ$ - excludes a pure-proton composition. At the energy range of $\sim100$ to $\sim200$~EeV, the proton horizon indeed becomes smaller than that of Fe, implying that Fe CRs would be more isotropic in this energy range. However, the results presented in their Fig.~7 for the number of pairs expected at different source densities and CR compositions in the absence of magnetic fields (which are largely consistent with the results of our simulations) do not actually rule out a pure-proton composition at a density $\sim 10^{-3}~\text{Mpc}^{-3}$ (the claim to the contrary in their abstract, and the constraints presented in their Fig.~11, are derived from comparing the proton simulations to isotropic simulations and thus do not constitute an observational constraint). Furthermore, the analysis presented does not include a treatment of uncertainties in observed CR energy estimates, which may have a dramatic effect at these high energies (a small shift in energy changes the observed ray count drastically), and does not include an estimate of the impact on the significance of the result due to the arbitrary choice of the value of the angular separation examined, $3^\circ$.

\citet{ding_imprint_2021} find that the observed dipole anisotropy is incompatible with a pure proton composition, which again seems to go against our results. Their main argument against a proton composition is the misalignment of the dipole direction measured in the data to the model. However, this holds only in the $8~\text{EeV}<E<32~\text{EeV}$ range, which we hadn't analyzed due to a lack of publicly available data; in the $E>32$~EeV energy bin, the directions actually do seem to be consistent at a 95\% CL. As discussed in the introduction, the implications of the lower-energy $\sim10$~EeV dipole remain uncertain.

We did not explicitly look for smaller-scale anisotropies in this work, but the mean arrival maps in Fig.~\ref{fig:meanmaps_iso} still roughly let us evaluate claims of smaller-scale excesses. In \citet{abdul_halim_large-scale_2024}, an excess in the Centaurus region of about $\sim1.1$ is found in Auger data after smearing by a radius of $45^\circ$, consistent with our map, since we get $\sim1.12$ for the Centaurus region after smearing the proton arrival map by the same angular radius. In \citet{kim_medium-scale_2025} TA analysis yields an excess in the Perseus-Pisces region. The actual value of the excess is not reported, but we estimate it to be $\sim1.6$ on a small angular scale, based on their figures. Our proton arrival map has $\sim1.5$ for Perseus-Pisces without smearing. The TA hotspot does not appear on our map, and indeed it is not correlated with any known structure. We chose not to analyze such small scales because of the possible strong coherent deflections for nuclei that would move the signal, but it just happens that both the Perseus-Pisces and Centaurus regions are not affected much by deflections in the UF23 models, so their detection in the data is not a strong point of evidence for either composition model. The Virgo cluster is the strongest anisotropy in our nuclei arrival map, and \citet{kim_medium-scale_2025} explicitly do not find any excess around it, but this could be either the result of a proton composition, strong deflections for nuclei, or cosmic variance, since the cluster is very close to Earth.

\section{Summary and conclusions}
\label{sec:discussion}

We have described a novel method for simulating UHECR arrival maps, including a semi-analytic method for propagating the spectrum of UHECR nuclei that is detailed in \S~\ref{sec:analytic_treatment}. The expression we derived for the relation between the source distance, source energy and observed energy, given in Eq.~(\ref{eq:dtilde_by_E}), matches simulations well as seen in Fig.~\ref{fig:econtours} and allows us to calculate important quantities like the rigidity distribution (Fig.~\ref{fig:rigidity}) and the source distance distribution (Fig.~\ref{fig:dndr}) of nuclei UHECRs.

We neglected EGMF deflections and chose our observables to be as insensitive as possible to the exact configuration of the GMF. We estimated the strength of random deflections both from measurements of the interstellar medium and using RM measurements and only looked at observables on large enough scales to not be affected by it. To handle the larger deflections by the ordered field, we either look at very large angular scales as is described in \S~\ref{sec:testlarge} or at general properties of the distribution like the entropy (\S~\ref{sec:testentropy}) and the correlation between different energy bins (\S~\ref{sec:testenergy}) that are not affected significantly by deflections. If the magnetic fields are considerably stronger than our estimates suggest, our results would be significantly affected, but we have taken care to be conservative where we can.

All of our results confirm that the data are highly isotropic. When applied to the available Auger Phase 1 data for rays with $32>~\text{EeV}$, the test of a large-scale correlation with the LSS discussed in \S~\ref{sec:results_corr} shows a very weak signal. To make models consistent with the data, the source density must be low enough that the "cosmic variance" is sufficient to mask the LSS structure. A source density of $s_0=10^{-2}~\text{Mpc}^{-3}$ is ruled out with a p-value of $<0.1\%$ even for the relatively isotropic proton model. For both proton and heavy-nuclei models, the p-value increases beyond $10\%$ (when accounting for possible systematic error in the Auger energy calibration) for source densities $s_0\le 10^{-4}~\text{Mpc}^{-3}$. For heavy nuclei models, both low source density and GMF deflections are required to account for the observed $T$ value with $s_0\approx 10^{-4}~\text{Mpc}^{-3}$, while $s_0\le 10^{-5}~\text{Mpc}^{-3}$ would be required in the absence of such deflections. In general, a heavy composition should yield a less isotropic distribution, but the large magnetic deflections of nuclei lower their LSS correlation signal to be comparable to protons.

While the large-scale correlations are mostly useful for constraining the source density, the results for the entropy, discussed in \S~\ref{sec:results_entropy}, also allow us to constrain the composition. This is because entropy is less affected by coherent magnetic deflections than the correlation statistic is. The high entropy of the available data implies a light composition with $\sim95\%$ confidence for a low, $10^{-4}~\text{Mpc}^{-3}$, source density, and $\sim99\%$ for a high, $10^{-2}~\text{Mpc}^{-3}$, source density. The physical reason for this difference between compositions is that nuclei must ultimately come from much closer distances (as seen in Fig.~\ref{fig:dndr}) and therefore from many fewer sources. This fact holds at least up to $60$~EeV, but at sufficiently high energies ($E\gtrsim100$~EeV), the proton horizon is sufficiently reduced for this method to become inefficient in discriminating between compositions. It's worth noting that this reasoning only applies to the specific heavy-nuclei models we are testing, which feature the narrow rigidity spectrum inferred from depth-of-shower-maximum measurements. In this sense, we are not constraining a heavy composition in general, but rather the "Auger mix" heavy composition models. Looking at Fig.~\ref{fig:dndr}, it's apparent that the iron component of the spectrum arrives from distances comparable to those of protons; hence, a pure iron composition, e.g., might be compatible with the observed isotropy.

Similar conclusions are drawn from the reconstructed dipole direction and amplitude as discussed in \S~\ref{sec:results_dipole}. The model predictions for the amplitude of the dipole at $E>32$~EeV are in tension with the observed dipole, which is consistent with isotropy, even for a low source density. Heavy nuclei models are inconsistent with the data at 99\% CL, while proton models are inconsistent with the data at lower CL, 94\% and 92\% for source densities of $10^{-2}~\text{Mpc}^{-3}$ and $10^{-4}~\text{Mpc}^{-3}$ respectively.  Using the full Auger data, not just the public dataset above $32$~EeV, might shed light on this issue, since our results agree with the observed dipole in a lower-energy bin.

It is interesting to compare our constraints on the source density to the results of \citet{pierre_auger_collaboration_bounds_2013}, which obtains a lower limit of roughly $10^{-4}~\text{Mpc}^{-3}$ from the two-point correlation function of the data (although this work does not take into account the effect of the ordered GMF). This number is essentially the upper limit we infer; taking both methods into account, the density is very tightly constrained.

\begin{figure*}
    \centering
    \includegraphics[width=1\linewidth]{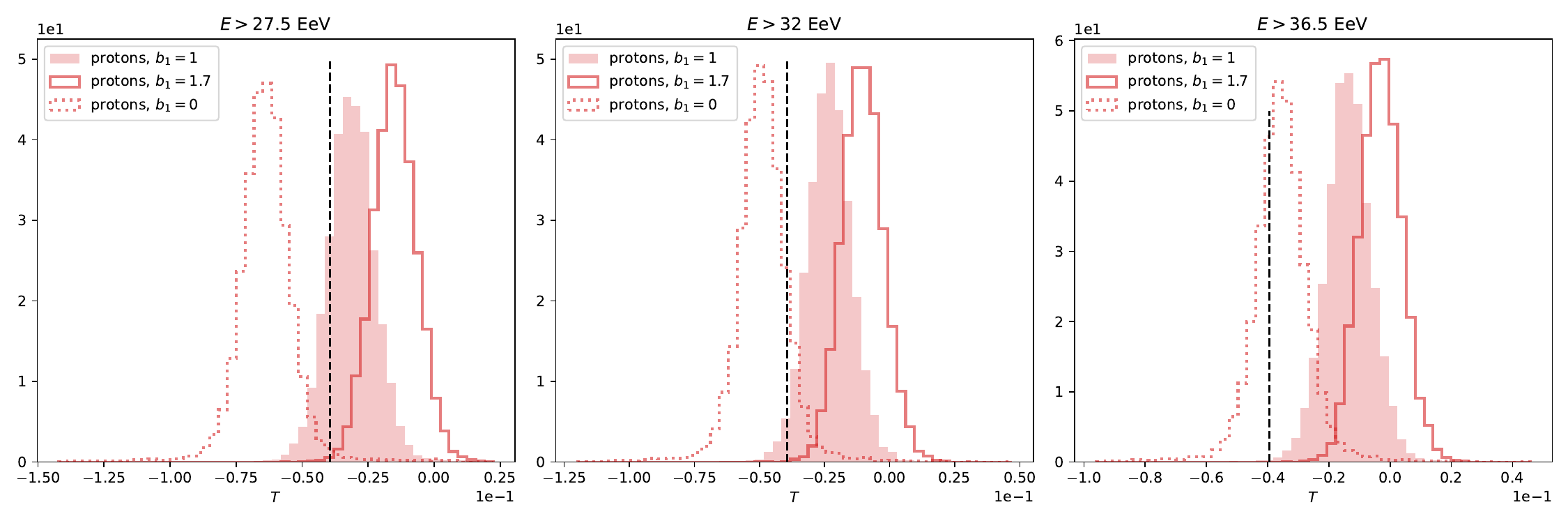}
    \caption{
    Correlation of proton arrival maps to the LSS, for an Auger-like exposure, $s_0=10^{-4}~\text{Mpc}^{-3}$, different source bias parameters, and for rays above different energy thresholds. The black dashed line is the value calculated from the available data, reported for $E>32$~EeV; the different energies shown in this plot represent the possible systematic error in Auger energy measurements. The statistic presented here as $T$ is similar to, but not exactly, the value described in \S~\ref{sec:testlarge}; instead of using few very large angular bins, here we divide the sky to smaller bins, a healpy \citep{zonca_healpy_2019} map with nside=4 (which has a resolution of $\sim15^\circ$), taking advantage of the small angular deflections of protons. The distribution of $T$ for an isotropic source distribution is well separated from that of a distribution correlated with the LSS, but the current uncertainty in the absolute energy calibration implies that the data cannot be used to determine which is preferred.}
    \label{fig:results_small_auger_calib}
\end{figure*}

If the source density is indeed as low as $10^{-4}~\text{Mpc}^{-3}$, so that the LSS signature is masked out by cosmic variance, increasing the experimental exposure and the number of detected cosmic-rays will in itself not enable one to identify the LSS signatures. Reducing the absolute energy calibration uncertainty may, on the other hand, allow the detection of the LSS correlation for proton models. Fig.~\ref{fig:results_small_auger_calib} demonstrates that the energy uncertainty has a noticeable effect on the results for protons, and given more precise energy measurements, it should be possible to detect the LSS signal for protons on smaller angular scales. As mentioned in the introduction, the reported correlation signals from some local structures (Centaurus for Auger, \citet{abdul_halim_large-scale_2024}, Perseus-Pisces for TA, \citet{kim_medium-scale_2025}) and with star-forming galaxies \citep{abreu_arrival_2022} suggest a proton composition.

The results of our analysis are in tension with the "Auger mix" heavy composition models inferred from the Auger depth-of-shower-maximum analyses \citep{pierre_auger_collaboration_depth_2026}. This intriguing tension remains unresolved. It should be noted in this context that TA shower depth measurements \citep{kim_recent_2026} show some discrepancies with Auger's results; while the $X_\text{max}$ measurements are generally compatible given current systematic and statistical uncertainties \citep{pierre_auger_collaboration_depth_2023}, some tensions between the two data sets remain, which are directly relevant to the inferred composition: (i) The Auger data show an elongation rate (the energy derivative of $X_{\text{max}}$) that is compatible with a proton composition in the energy range of $10^{18}-10^{18.5}$~eV and a lower elongation rate, that may be interpreted as a transition to heavier composition, at higher energy, while the TA data \citep[similar to earlier Fly's Eye data, e.g.,][]{pierog_air_2017} show a larger elongation rate, consistent with a decreasing CR primary mass, at $10^{18}-10^{18.5}$~eV (and compatible with protons at higher energy); (ii) The decrease in $\sigma(X_{\text{max}})$ observed in Auger data, suggesting a transition to heavier composition, above $10^{19}$~eV is not observed in the TA data (note that the variable aerosol contribution, which is not included in the TA analysis, provides only a small contribution to the inferred $\sigma(X_{\text{max}})$, which does not explain the difference between the data sets; see, e.g., \citet{bergman_testing_2023}). These tensions are under ongoing investigation by the joint Auger-TA working group. Notably, differences between different experiments' composition measurements exist also at energies well below the ankle, making it difficult to anchor the Auger and TA results to other experiments at lower energies; see, for example, the large differences between the Fe fractions inferred by the KASCADE-Grande, IceCube/IceTop and Auger experiments at $\sim10^{17.5}$~eV \citep{kang_recent_2023}.

Another possible challenge to a pure proton model of UHECRs is constraints from non-detection of high-energy neutrinos. The IceCube collaboration analysis \citep{icecube_collaboration_search_2025} constrains the proton fraction above 30~EeV to be $\le70\%$ at 90\%~CL assuming source evolution similar to or stronger than the star formation rate. The Auger upper limit for neutrinos is slightly lower \citep{alvarez-muniz_searches_2025, pierre_auger_collaboration_investigating_2023} and their constraints for protons are similar. These are not yet stringent constraints on the proton fraction, since even a pure proton model with strong evolution is not ruled out at a 99\%~CL.

Finally, we note that in this work, we have assumed the sources to be steady. If the sources are transient, the number of sources contributing to the UHECR flux should decrease more rapidly with energy, and different sources may shine at different energies \citep{waxman_images_1996}. A detailed discussion of transient sources is beyond the scope of the current paper.

\begin{acknowledgments}
We wish to thank Foteini Oikonomou and Michael Unger for their useful discussions with us, and Karl-Heinz Kampert, Teresa Bister, Kohta Murase, Esteban Roulet, Alexander Korochkin, Dmitri Semikoz, Peter Tinyakov and Mikhail Kuznetsov for their helpful comments. This research was partially supported by ISF, Minerva, and Segre grants.

\end{acknowledgments}

\end{document}